\documentclass[3p,times,twocolumn]{elsarticle}
\usepackage{amssymb}
\usepackage{amsthm}
\usepackage{lineno}
\usepackage{subfigure}
\usepackage{xspace}
\newcommand{\sigttbar  }{\ensuremath{160~\mathrm{pb}}\xspace}
\newcommand{\dsigttbar }{\ensuremath{10\%}\xspace}
\newcommand{\ttbcross}{\ensuremath{\sigma_\mathrm{t\bar{t}}}\xspace}
\newcommand{\ttbcrossv}[1]{\ensuremath{\sigma_\mathrm{t\bar{t}}(\mathrm#1)}\xspace}
\newcommand{\ttbar   }{\ensuremath{\mathrm{t\bar{t}}}\xspace}

\newcommand{\ttbarlj }{\ensuremath{\ttbar\to\mbox{lepton+jets}}\xspace}

\newcommand{\WB      }{\ensuremath{\mbox{W boson}}\xspace}
\newcommand{\ZB      }{\ensuremath{\mbox{Z boson}}\xspace}

\newcommand{\arap    }{\ensuremath{\vert y\vert}\xspace}
\newcommand{\arapm   }{\ensuremath{\vert y\vert_\mathrm{max}}\xspace}
\newcommand{\aeta    }{\ensuremath{\vert\eta\vert}\xspace}
\newcommand{\Mjj     }{\ensuremath{M_{JJ}}\xspace}
\newcommand{\rts     }{\ensuremath{\sqrt{s}}\xspace}
\newcommand{\gev     }{\ensuremath{\mathrm{GeV}}\xspace}
\newcommand{\GeV     }{\ensuremath{\gev}\xspace}
\newcommand{\tev     }{\ensuremath{\mathrm{TeV}}\xspace}
\newcommand{\TeV     }{\ensuremath{\tev}\xspace}
\newcommand{\pt      }{\ensuremath{p_{\mathrm{T}}}\xspace}
\newcommand{\ptq     }{\ensuremath{p^{2}_{\mathrm{T}}}\xspace}
\newcommand{\ptb     }{\ensuremath{\overline{p}_{\mathrm{T}}}\xspace}
\newcommand{\Ord     }[1]{\ensuremath{{\mathcal{O}}(#1)}\xspace}
\newcommand{\qn      }{\ensuremath{Q_0}\xspace}
\newcommand{\HT      }{\ensuremath{H_{\mathrm{T}}}\xspace}
\newcommand{\kt      }{\ensuremath{k_\mathrm{t}}\xspace}
\newcommand{\XZ      }[3]{\ensuremath{#1\,\pm #2_\mathrm{stat}\,\pm #3_\mathrm{syst}}\xspace}
\newcommand{\Y       }[3]{\ensuremath{#1\,^{+\,#2}_{-\,#3}}\xspace}
\newcommand{\mt      }{\ensuremath{m_{\mathrm{top}}}\xspace}
\newcommand{\msb     }{\ensuremath{{\scriptstyle\overline{\rm MS}}}\xspace}
\newcommand{\mtmsb   }{\ensuremath{m_{\mathrm{top}}^{\mathrm{\msb}}}\xspace}
\newcommand{\mtpole  }{\ensuremath{m_{\mathrm{top}}^{\mathrm{pole}}}\xspace}
\newcommand{\mtMC    }{\ensuremath{m_{\mathrm{top}}^{\mathrm{MC}}}\xspace}
\newcommand{\xk      }{\ensuremath{x_\mathrm{k}}\xspace}
\newcommand{\xo      }{\ensuremath{x_\mathrm{1}}\xspace}
\newcommand{\xt      }{\ensuremath{x_\mathrm{2}}\xspace}
\newcommand{\Dy      }{\ensuremath{\Delta y}\xspace}
\newcommand{\Njets   }{\ensuremath{N_\mathrm{jets}}\xspace}
\newcommand{\RtW     }{\ensuremath{R_\mathrm{32}}\xspace}
\newcommand{\ejets   }{\ensuremath{\mbox{e+jets}}\xspace}

\newcommand{\mjets   }{\ensuremath{\mu\mbox{+jets}}\xspace}
\newcommand{\bquark  }{\ensuremath{b\mbox{-quark}}\xspace}
\newcommand{\bjet    }{\ensuremath{b\mbox{-jet}}\xspace}
\newcommand{\bjets   }{\ensuremath{b\mbox{-jets}}\xspace}
\newcommand{\btagged }{\ensuremath{b\mbox{-tagged}}\xspace}
\newcommand{\al      }{\ensuremath{\alpha_{s}}\xspace}
\newcommand{\mzq     }{\ensuremath{M^2_\mathrm{Z}}\xspace}
\newcommand{\almzq   }{\ensuremath{\al(\mzq)}\xspace}
\newcommand{\mttevo  }{\ensuremath{\XZ{173.18}{0.56}{0.75}}\xspace}
\newcommand{\Alpgen  }{{\sc Alpgen}\xspace}
\newcommand{\Herwig  }{{\sc Herwig}\xspace}
\newcommand{\Hej     }{{\sc HEJ}\xspace}
\newcommand{\Jimmy   }{{\sc Jimmy}\xspace}
\newcommand{\Madgraph}{{\sc MadGraph}\xspace}
\newcommand{\Mcatnlo }{{\sc MC@NLO}\xspace}
\newcommand{\Mcfm    }{{\sc MCFM}\xspace}
\newcommand{\Nlojet  }{{\sc NLOJet++}\xspace}
\newcommand{\Powheg  }{{\sc Powheg}\xspace}
\newcommand{\Pythia  }{{\sc Pythia}\xspace}
\newcommand{\Sherpa  }{{\sc Sherpa}\xspace}
\DeclareGraphicsExtensions{.pdf,.png}
\begin{document}
\begin{frontmatter}
%
%
\title{QCD results from the LHC}
\author{Richard Nisius\fnref{label1}}
\fntext[label1]{Invited talk given at the {\it Ringberg Workshop: New Trends in
    HERA Physics 2011, Ringberg Castle, Germany, 25-28 September, 2011},
  Email: Richard.Nisius@mpp.mpg.de}
\address{Max-Planck-Institut f\"ur Physik, F\"ohringer Ring 6, 
         D-80805 M\"unchen, Germany}
%
\begin{abstract}
 A selection of results from the 2010 data taking period of the ATLAS and CMS
 experiments at the LHC at a proton-proton centre-of-mass energy of
 $\rts=7$~\TeV\ is presented.
 These results comprise differential jet cross sections for varying jet
 multiplicities, the investigation of properties of large rapidity gaps spanned
 by a dijet system, the production of heavy gauge bosons together with jets, and
 finally the investigations of properties of top quark pair production.
\end{abstract}
%
%
\end{frontmatter}
%
%
\section{Introduction}
\label{sec:intro}
 The LHC is a wonderful QCD machine. The large proton energy allows for probing
 predictions of QCD at unprecedented energy scales in accelerator physics.
 Due to the high luminosity of the machine, and large QCD cross sections,
 especially for jet production, many analyses are limited by systematic
 uncertainties right from the start.
 This puts high demands on the understanding of the detectors, and also calls
 for high performance jet algorithms to cope with the ever increasing
 complications due to complex final states, and the occurrence of more than one
 proton-proton interaction per bunch crossing (pileup) that potentially
 deteriorates the jet energy resolution.

 The results presented are all based on the LHC running at a proton-proton
 centre-of-mass energy of $\rts=7$~\TeV.
 For many of the topics discussed, results from the ATLAS and CMS experiments
 exist, however, due to the limited space, for each topic only a single result
 is shown, concentrating on published results, i.e.~additional preliminary
 measurements are not included.

 The paper is organised as follows: jet production for increasing jet
 multiplicities is discussed in Section~\ref{sec:jetpro}. Adding a further hard
 scale, heavy gauge boson plus jet production is detailed in
 Section~\ref{sec:wzboson}.
 A number of issues related to the production of top quarks are highlighted
 in Section~\ref{sec:ttbar}.
 Finally, the summary and conclusions are given in Section~\ref{sec:concl}.
%
%
\section{Jet production}
\label{sec:jetpro}
\begin{figure*}[tbph!]
\centering
\includegraphics[width=0.98\textwidth]{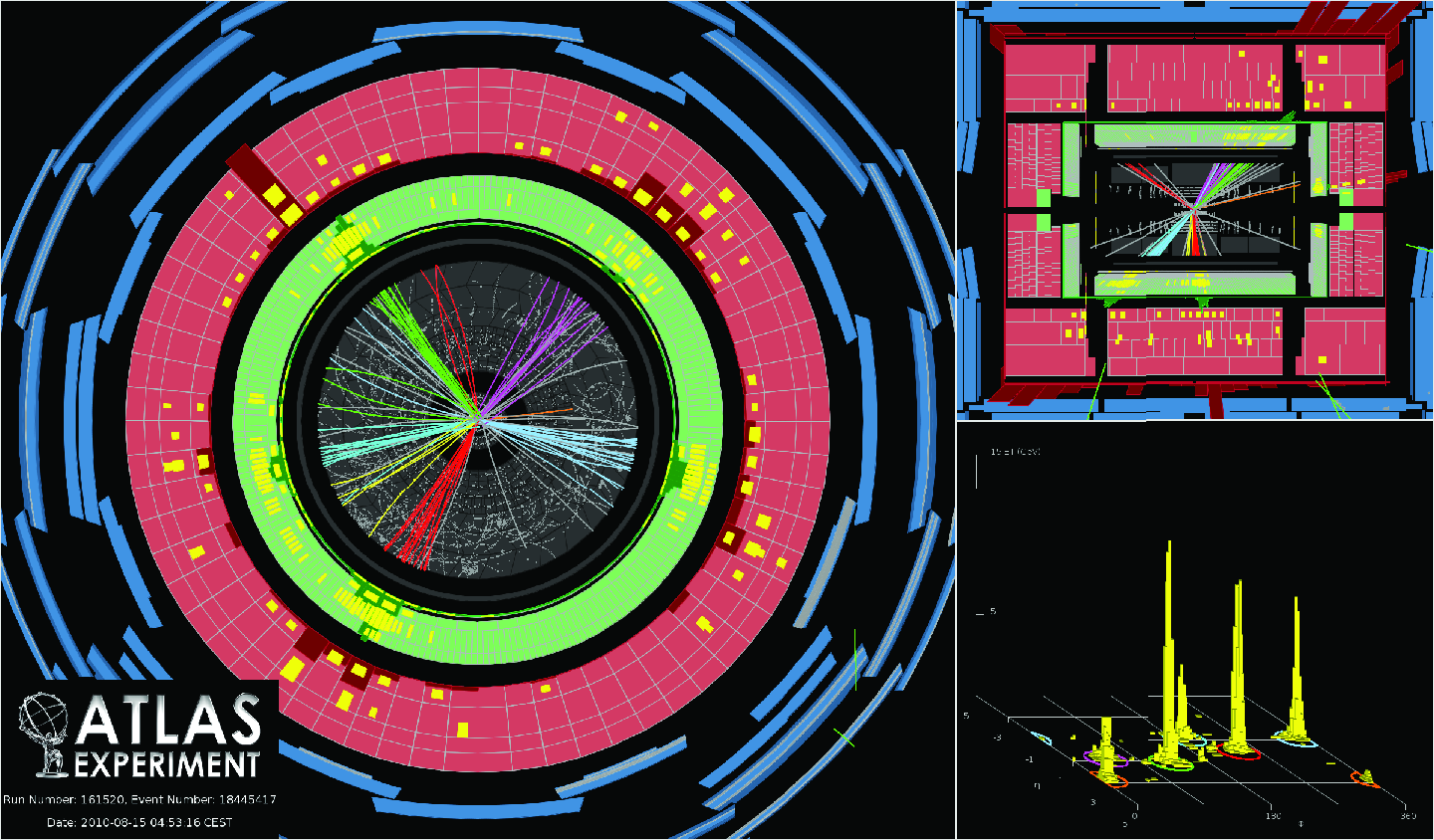}
\caption{Event display of a six-jet event in the ATLAS
  detector~\protect\cite{ATL-2011-030}. Shown are: a view along, and a view
  parallel to the beam axis, and the angular distribution of transverse energy
  in pseudo rapidity ($\eta$) and azimuthal angle ($\Phi$). The measured jet
  transverse momenta range from $\pt=84$~\GeV to $\pt=203$~\GeV.
  \label{fig:sixjet}}
\end{figure*}

 Due to the large proton energies, and the correspondingly large phase space for
 jet production, very complex final states, with large jet transverse momenta
 (\pt) occur.
 An example of this, a six jet event observed in the ATLAS detector, is shown in
 Figure~\ref{fig:sixjet}.
 To properly reconstruct those final states, a high performance jet algorithm is
 needed. The present choice of the LHC experiments for this is the anti-\kt\
 algorithm~\cite{CAC-0801,CAC-0802}, which is a sequential clustering algorithm
 that uses 1/\ptq\ as the weighting factor for the scaled distance, and the
 $R$ parameter to define the jet resolution.
 This algorithm exhibits a number of good features~\cite{CAC-0801}, which, on
 top of its infra-red safety, makes it superior to other possible choices.
%
%
 These features are: round and rigid jet shapes that lead to a clear and stable
 spatial definition of jets; an almost \pt\ independent jet area, ensuring an
 almost constant pileup correction as a function of \pt; and finally, very small
 back reaction, i.e.~re-assignments of particles from the hard interaction to
 jets for different pileup contributions, which guarantees stable definitions of
 the jets stemming from the hard QCD process.
 In search for optimised uncertainties of observables the $R$ parameter is
 varied in the comparisons to the theoretical predictions, with typical values
 in the range $R=0.4-0.7$. In this respect, jets with smaller R values are found
 to be less dependent on pileup, and those with larger R values to be less
 dependent on scale changes in the theoretical predictions~\cite{ATL-2011-030}.

 For the ATLAS and CMS analyses, the observed jet distributions are corrected
 for detector effects, and then, at the resulting stable particle level,
 compared to the theoretical predictions which come in a large number of
 flavours.
 These comprise: leading-order (LO) $2\to2$ Matrix Elements~(ME) plus subsequent
 Parton Shower~(PS) and underlying event~(UE) implemented in the programs
 \Pythia~\cite{SJO-0601} and \Herwig~\cite{COR-0001} together with
 \Jimmy~\cite{BUT-9601} (those will be referred to as LO $2\to2$ predictions);
 LO $2\to n$ MEs provided by the \Sherpa~\cite{GLE-0901},
 \Madgraph~\cite{STE-9401,ALW-0701} and \Alpgen~\cite{MAN-0301} programs, with
 subsequent internal (\Sherpa) or external, i.e.~by other packages provided,
 (\Madgraph, \Alpgen) PS and UE (referred to as LO $2\to n$ predictions);
 NLO ME calculations for $n\le 3$ outgoing partons featured by the
 \Mcfm~\cite{CAM-0301} and \Nlojet~\cite{NAG-0301} programs;
 NLO ME calculations matched to PS that are either provided by the
 \Mcatnlo~\cite{FRI-0201,FRI-0301} together with the \Herwig\ software packages,
 or by the \Powheg~\cite{FRI-0701} generator interfaced to either the
 \Pythia\ or \Herwig\ programs;
 and finally, an all order resummed calculation for wide angle emissions
 implemented in the \Hej~\cite{AND-1001,AND-1101} program.
 In addition, for the soft parts of the event simulation, a number of different
 tunings of parameters that control those parts in the general purpose Monte
 Carlo programs are in use, see for example~\cite{SKA-1001}.
 This makes up for a very large variety of predictions for comparisons, a
 number of them are shown below.
%
\begin{figure}[tbp!]
\centering 
\includegraphics[width=0.48\textwidth]{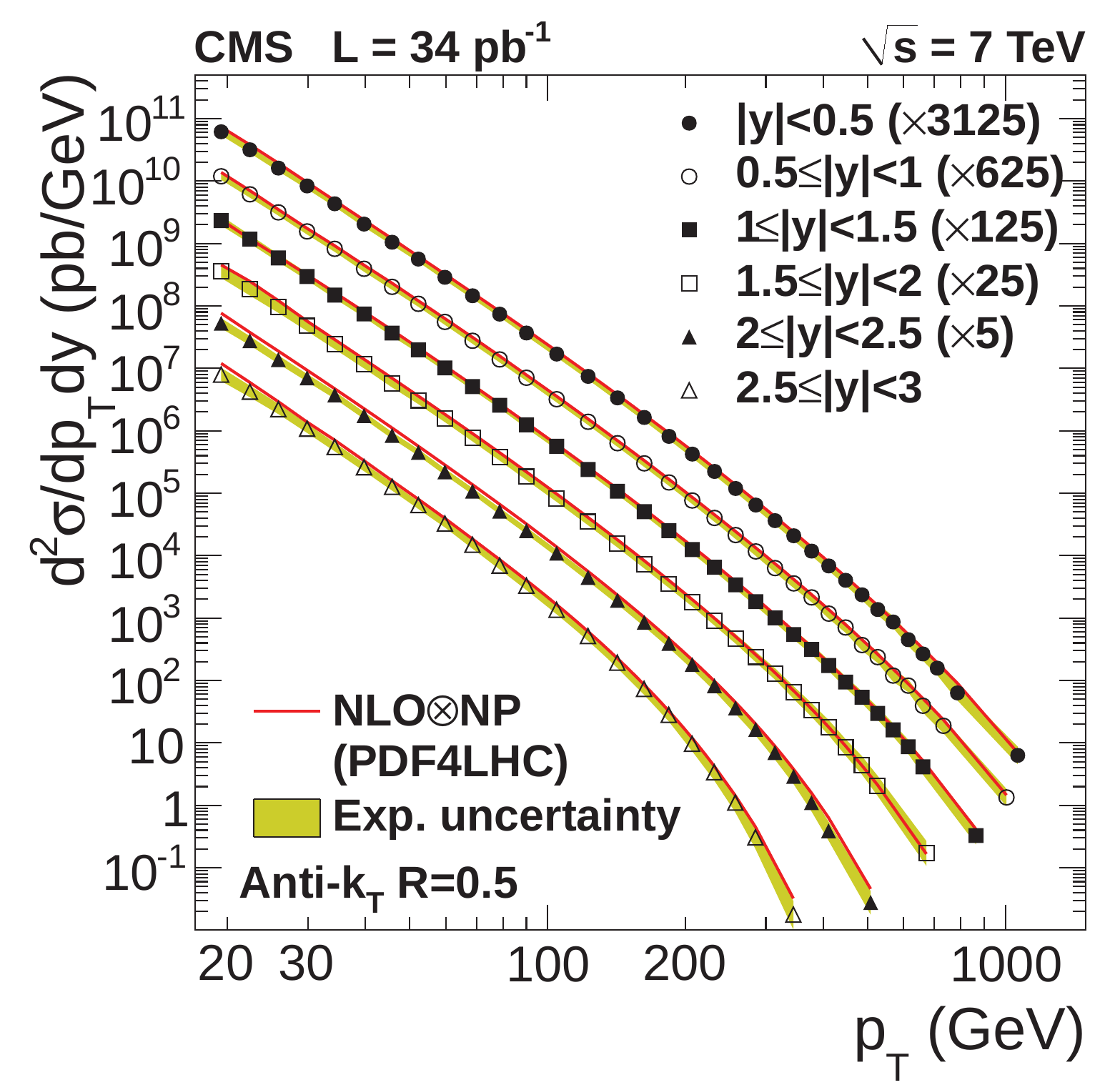}
\caption{Inclusive jet production~\protect\cite{CMS-1104}. Shown is the double
  differential inclusive jet cross section as a function of \pt, and in bins
  of the absolute rapidity.
  \label{fig:onejet}}
\end{figure}
%

 In Figure~\ref{fig:onejet} the double differential inclusive jet
 cross section~\cite{CMS-1104} as a function of \pt, for the range $18~\GeV< \pt
 <1100~\GeV$, and in bins of the absolute value of the rapidity \arap, is shown.
 Already now, the reach in \pt\ scales that are probed by the LHC experiments
 extend those probed by the Tevatron experiments from about 700~\GeV to about
 1100~\GeV.
 The data are compared to the NLO prediction from the \Nlojet\ program, with
 non-perturbative corrections estimated using the \Pythia\ and \Herwig\ models.

\begin{figure}[tbp!]
\centering 
\includegraphics[width=0.47\textwidth]{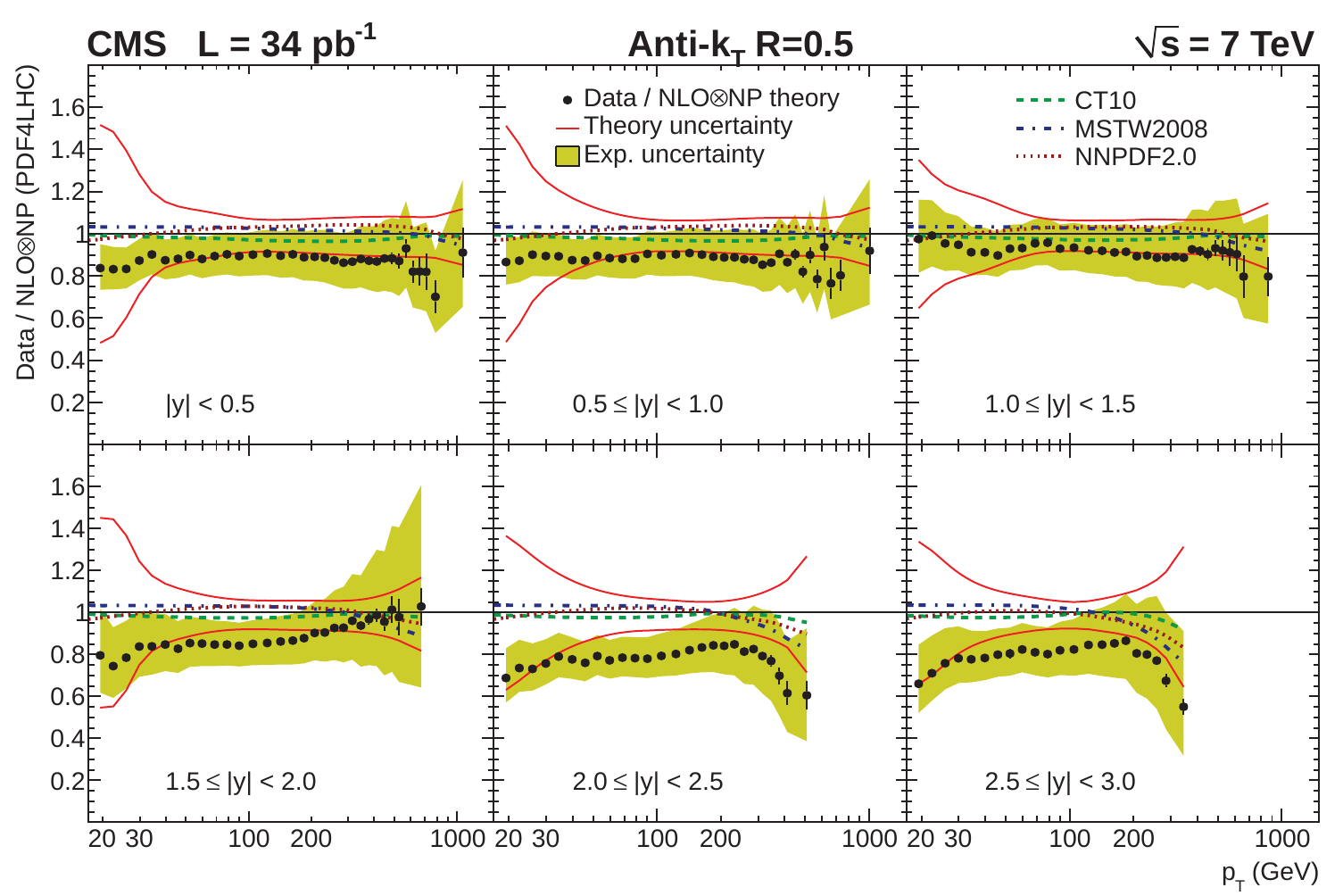}
\caption{Inclusive jet production~\protect\cite{CMS-1104}. Shown is the ratio of
  the measured cross section and the NLO prediction as a function of \pt, and in
  bins of the absolute rapidity.
  \label{fig:onejetrat}}
\end{figure}

 For CMS, jets are reconstructed using the so-called particle flow algorithm.
 This algorithm combines information from a list of objects: leptons, photons,
 and charged and neutral hadrons, into jets. For each of these objects, this
 information is obtained from various components of the detector.
 As for most of the inclusive jet cross section determinations, the experimental
 uncertainty is dominated by the Jet Energy Scale (JES) uncertainty.
 The description of the data by the NLO prediction over a large range in \pt\ is
 fair. However, the prediction is systematically higher than the data,
 especially so at large values of \arap, see Figure~\ref{fig:onejetrat}.
 At the highest \pt\ the theoretical uncertainty (shown as solid lines above and
 below unity) is dominated by the one from the Parton Density Functions (PDFs)
 of the proton, parametrised as a function of the partons momentum \xk\ from the
 proton, and consequently, the data start to constrain the PDFs, see
 Figure~\ref{fig:onejetrat}.
%
%

\begin{figure}[tbp!]
\centering
\includegraphics[width=0.47\textwidth]{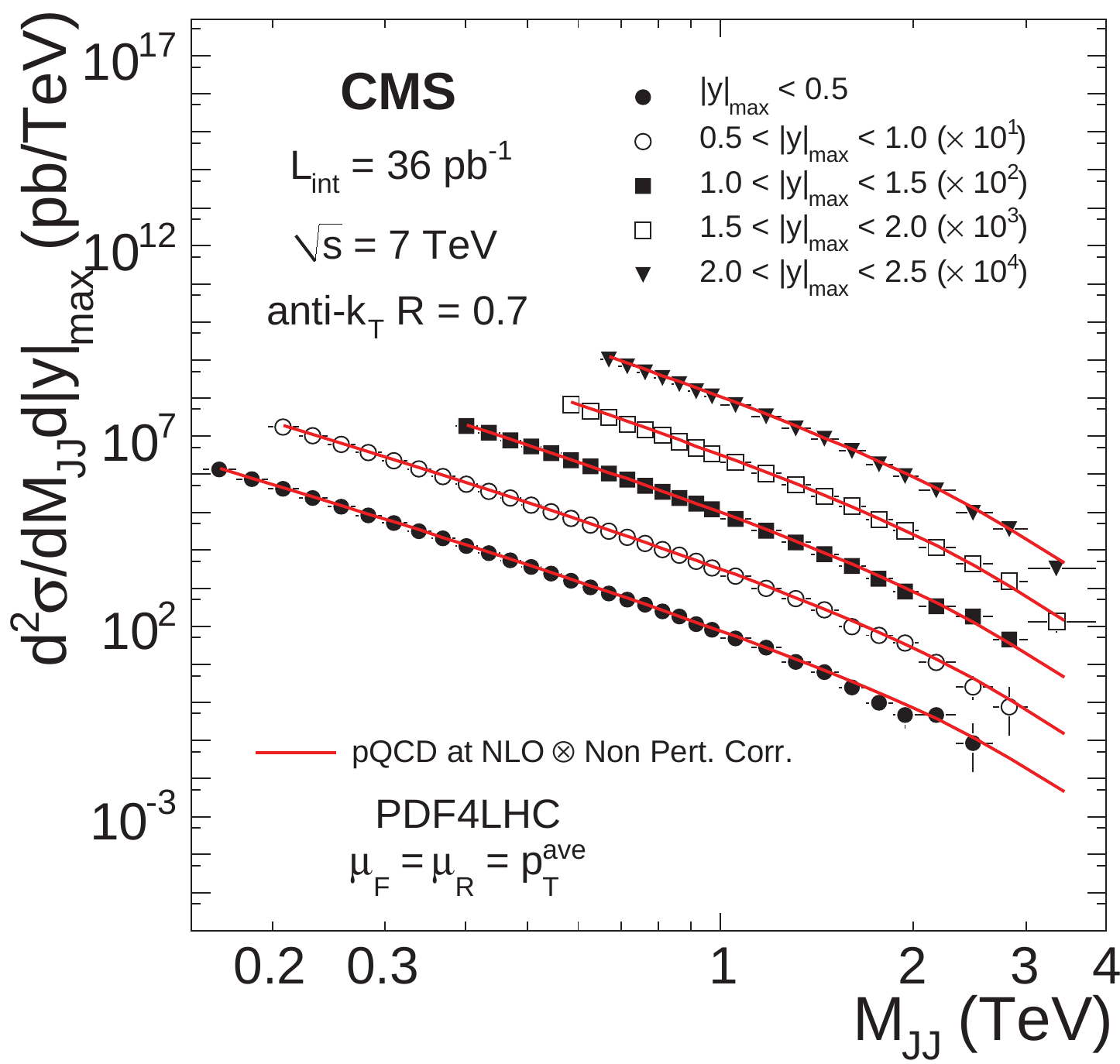}
\caption{Inclusive dijet production as a function of the dijet invariant mass
  \Mjj, and in bins of maximum of the absolute rapidity of the two jets
  \arapm~\protect\cite{CMS-1105}. The horizontal bars indicate the bin width,
  the vertical bars the statistical uncertainty.
  \label{fig:twojet}}
\end{figure}

\begin{figure}[tbp!]
\centering
\includegraphics[width=0.48\textwidth]{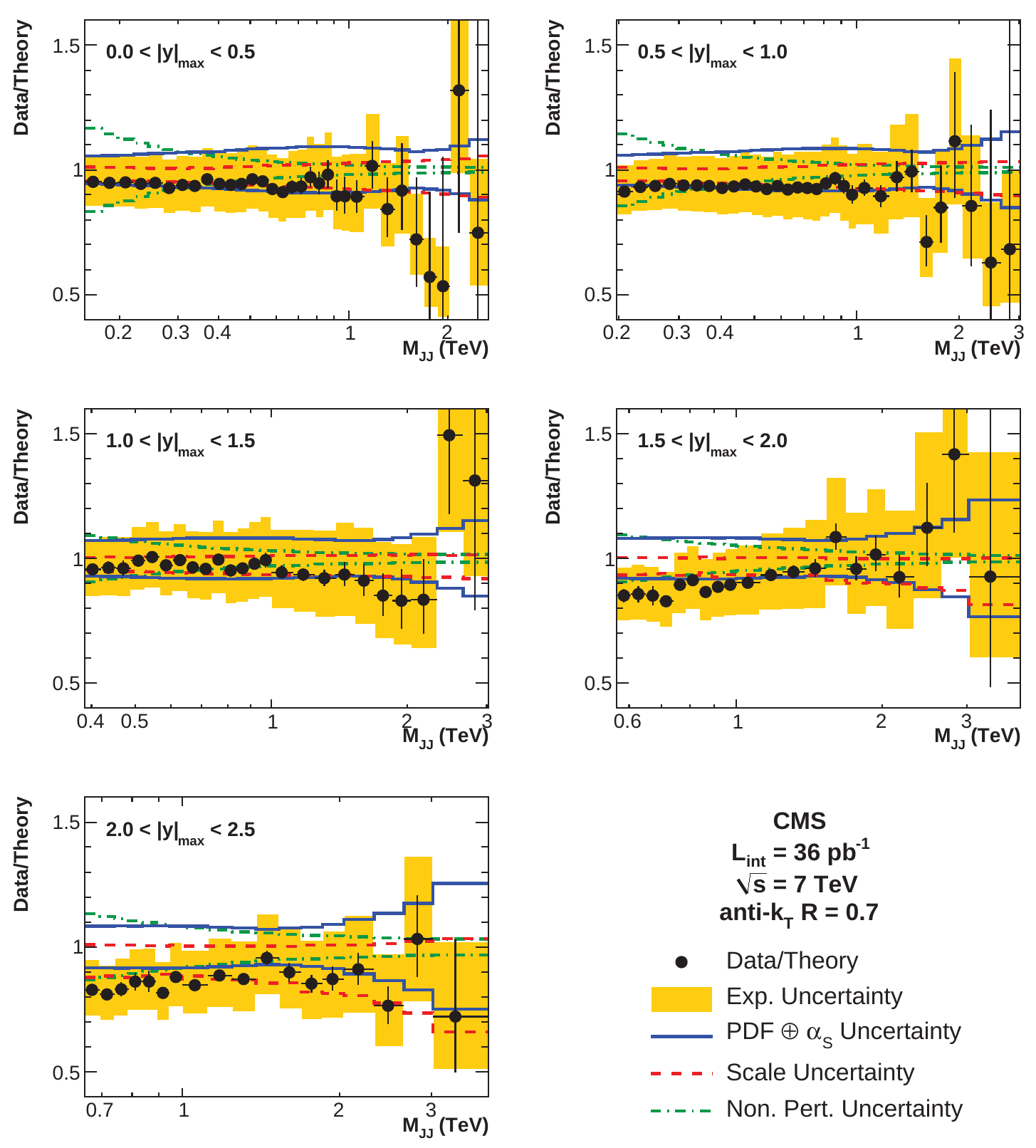}
\caption{Inclusive dijet production as a function of the dijet invariant mass
  \Mjj, and in bins of maximum of the absolute rapidity of the two jets
  \arapm~\protect\cite{CMS-1105}. Shown is the ratio of the observed and
  predicted cross sections. The data are displayed as in
  Figure~\ref{fig:twojet}.  The uncertainties of the prediction are shown as
  solid lines above and below unity.
  \label{fig:twojetrat}}
\end{figure}

 A similar conclusion holds for the double differential inclusive dijet
 production cross section shown in Figure~\ref{fig:twojet} as a function of
 \Mjj, the invariant mass of the two highest \pt\ jets in the event, and for
 various bins in \arapm, the maximum of the absolute rapidity of these two jets.
 In this analysis, the dijet production cross section probes the PDFs in the
 range $0.0008<\xo\xt<0.25$.
 The observable \Mjj\ can be reconstructed with a 7$\%$ (3$\%$) resolution at
 $\Mjj=0.2$~\TeV (3~\TeV).
 The low-mass region is mainly sensitive to the UE, whereas the high-mass region
 potentially constrains the high-x region of the PDFs, see
 Figure~\ref{fig:twojetrat}.
 At present, the component of the cross section uncertainty caused by the JES
 uncertainty amounts to 15$\%$ (60$\%$) at $\Mjj=0.2$~\TeV (3~\TeV).
%
%
 The cross section uncertainty due to the non-perturbative corrections mentioned
 above is comparably small, and amounts to 15$\%$ (2$\%$) at the same invariant
 masses.
 Finally, the present PDF uncertainty of the theoretical prediction is 5$\%$
 (30$\%$), again at the same invariant masses.
 Consequently, still some improvement in the JES uncertainty is needed to get a
 precise constraint for the high-x region of the PDFs.

\begin{figure}[tbp!]
\centering
\includegraphics[width=0.47\textwidth]{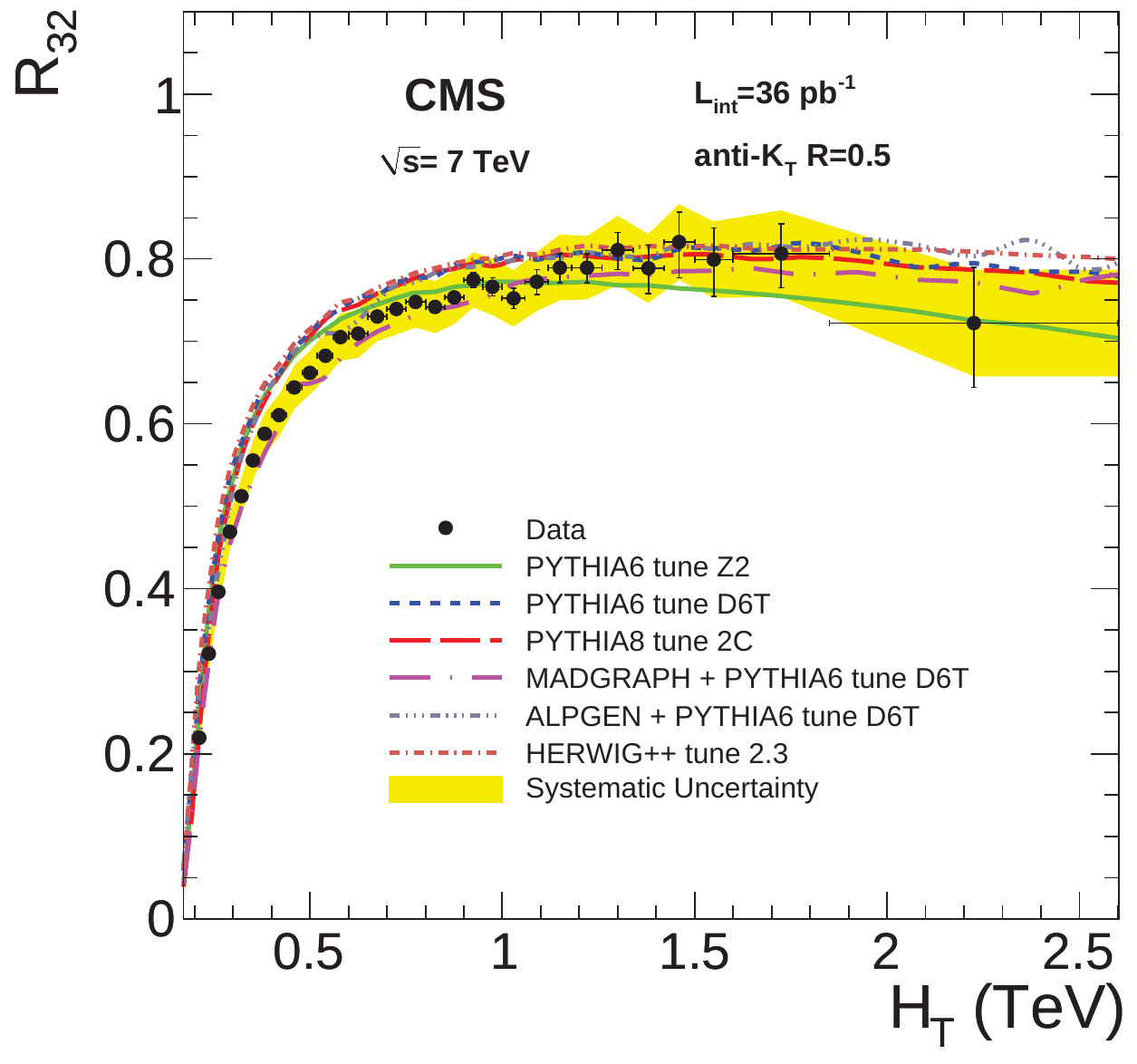}
\caption{The ratio of the 3-jet to 2-jet cross sections, \RtW, as a function of
  the scalar transverse momentum sum \HT~\protect\cite{CMS-1106}. The vertical
  bars denote the statistical uncertainties, the shaded area indicates the
  systematic uncertainty. The data are compared to a number of predictions
  explained in the text.
  \label{fig:ratjet}}
\end{figure}

 The next observable discussed is \RtW, the ratio of the 3-jet to 2-jet
 cross sections.
 For this analysis jets within $\arap=2.5$ and for $\pt>50$~\GeV\ are used.
 Being a ratio, \RtW, shown in Figure~\ref{fig:ratjet} as a function of \HT, the
 scalar sum of the \pt\ of all jets defined above, profits from cancellations of
 many systematic uncertainties.
 These are most notably the uncertainties from the imperfect knowledge of the
 jet energy scale and the one from the jet selection efficiency.
%
%
 In this analysis the efficiency for 2-jet events is 100$\%$, whereas the
 efficiency for 3-jet events increases from 72$\%$ at $\HT=0.2$~\TeV to 100$\%$
 at $\HT=0.4$~\TeV.
 The present resolution in \pt\ translates into a resolution in \HT\ of about
 6$\%$ (3.5$\%$) at $\HT=0.05$~\TeV (1~\TeV).
 The size of the correction to the particle level is small and only amounts to
 about 4$\%$ (2$\%$) for the two regions $\HT<0.5$~\TeV~($>0.5$~\TeV).
%
%

\begin{figure}[tbp!]
\centering
\includegraphics[width=0.48\textwidth]{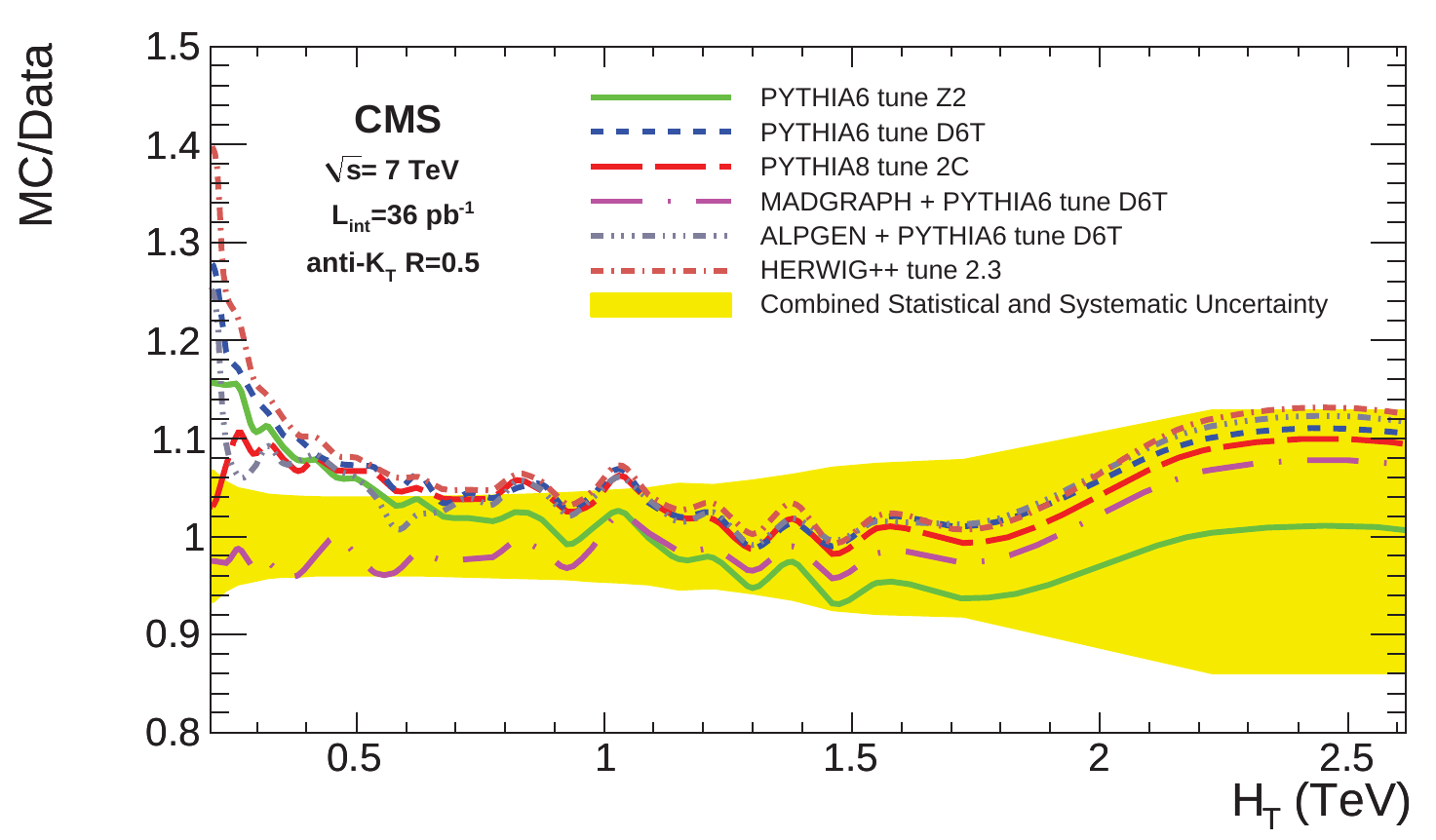}
\caption{The ratio of the 3-jet to 2-jet cross sections as a function of the
  scalar transverse momentum sum \HT~\protect\cite{CMS-1106}.  Shown is the
  ratio of the predicted and observed \RtW. The shaded area indicates the total
  experimental uncertainty.
  \label{fig:ratjetrat}}
\end{figure}

 In this ratio, the total experimental uncertainty amounts to (4-10)$\%$ and is
 mainly limited by the knowledge of the \pt\ dependence in the Monte Carlo
 predictions entering the correction procedure. Consequently, the \RtW\ ratio is
 considerably more precise than the inclusive jet cross sections discussed
 above.
 The ratio reaches a plateau of about 0.8, where the actual value depends on the
 choice of jet algorithm and the jet selection criteria like the chosen rapidity
 range.
 The plateau region can be nicely described by a number of predictions, see
 Figure~\ref{fig:ratjetrat}. However, the steep rise for low values of \HT,
 originating from the increasing phase space for the emission of the third jet,
 is only adequately followed by the \Madgraph\ prediction, i.e.~all other
 predictions are too steep.
%
%

\begin{figure}[tbp!]
\centering
\includegraphics[width=0.50\textwidth]{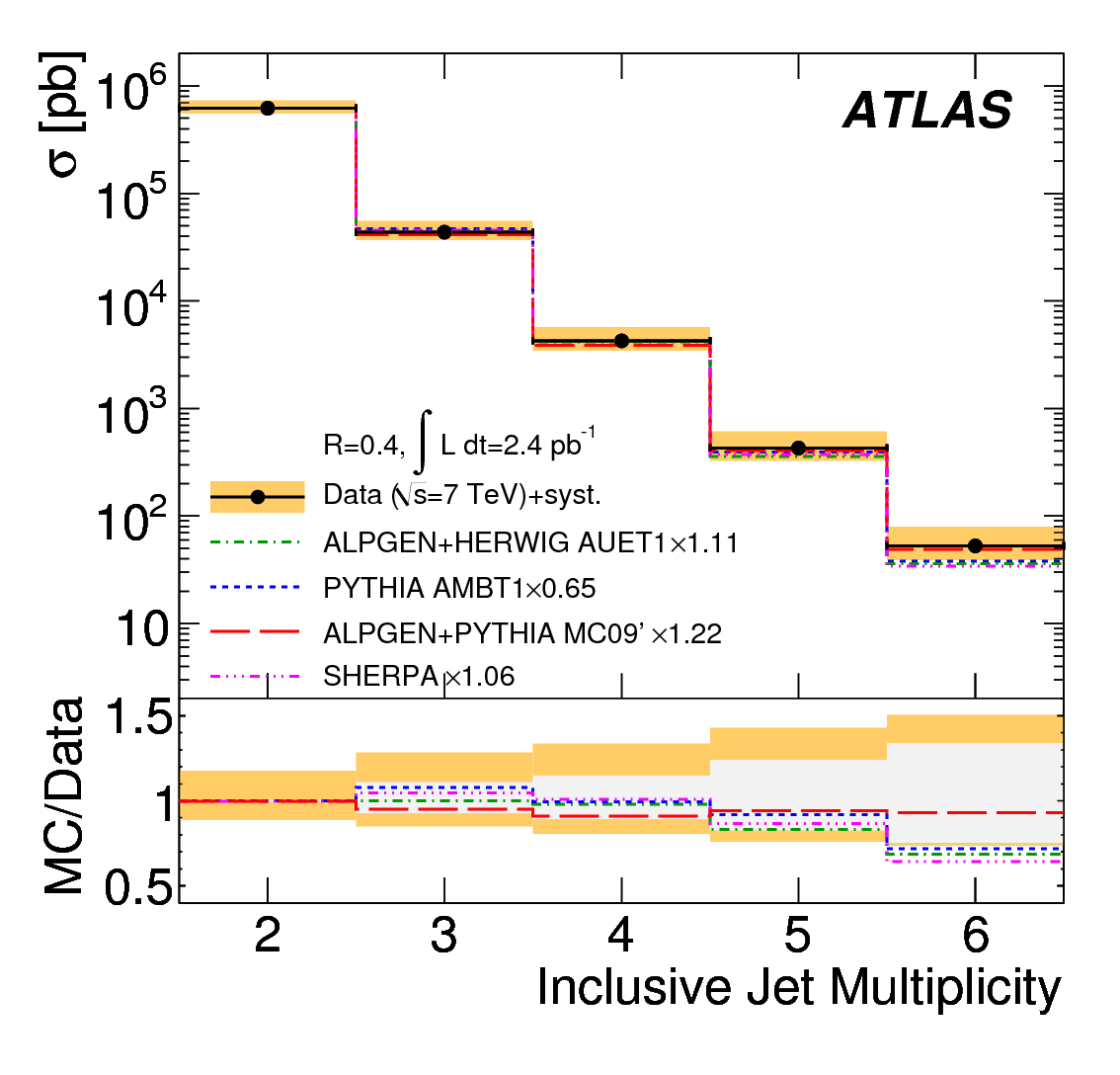}
\caption{Inclusive multijet cross section as a function of the jet
  multiplicity~\protect\cite{ATL-2011-030}. The darker shaded band corresponds
  to the systematic uncertainty excluding the contribution from the
  luminosity. The lighter shaded band corresponds to the systematic uncertainty
  on the shape of the measured distribution. The theoretical predictions are
  individually normalised to the $\Njets\ge2$ cross section.
  \label{fig:muljetincl}}
\end{figure}

 Finally, the cross section for multijet production for a minimum number of
 jets \Njets\ with $\Njets\ge2, \ldots 6$ is shown in
 Figure~\ref{fig:muljetincl}.
 Jets within $\arap=2.8$ are used, where in a \pt\ ordered list the first jet is
 required to fulfill $\pt>80$~\GeV, and all others to fulfill $\pt>60$~\GeV.
 The present JES uncertainty is asymmetric. It amounts to 5$\%$ (2.5$\%$) at
 $\pt=0.06$~\TeV (1~\TeV) and is larger than -3$\%$ everywhere.

 Within uncertainties, the shape of the inclusive jet multiplicity can be
 accounted for by all predictions~\cite{ATL-2011-030}, which however do show a
 slightly steeper trend than the data.
 There are very significant differences in the absolute predictions that result
 in different overall scaling factors, ranging from 0.65 up to 1.22, which are
 applied to individually normalise the predictions to the $\Njets\ge2$ bin, see
 Figure~\ref{fig:muljetincl}.
 The smallest scaling with only +6$\%$ is needed for the LO \Sherpa\ $2\to n$
 prediction, the largest with -35$\%$ for the LO \Pythia\ $2\to 2$ prediction.
%
%

\begin{figure}[tbp!]
\centering
\includegraphics[width=0.50\textwidth]{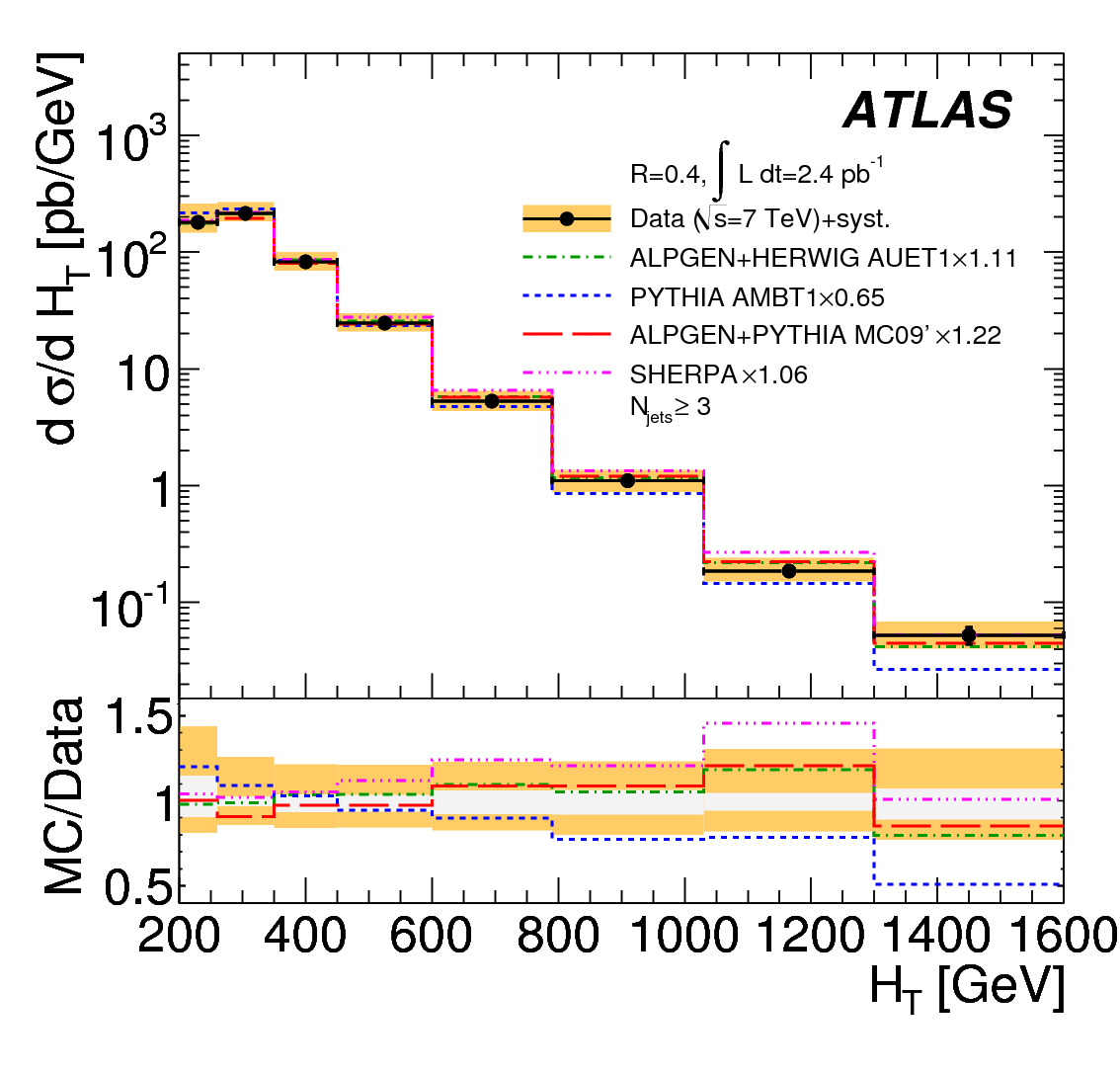}
\caption{Differential multijet cross section as a function of
  \HT~\protect\cite{ATL-2011-030}, for $\Njets\ge3$. The predictions are
  normalised as in Figure~\protect\ref{fig:muljetincl}.
  \label{fig:muljetpttj}}
\end{figure}

\begin{figure}[tbp!]
\centering
\includegraphics[width=0.50\textwidth]{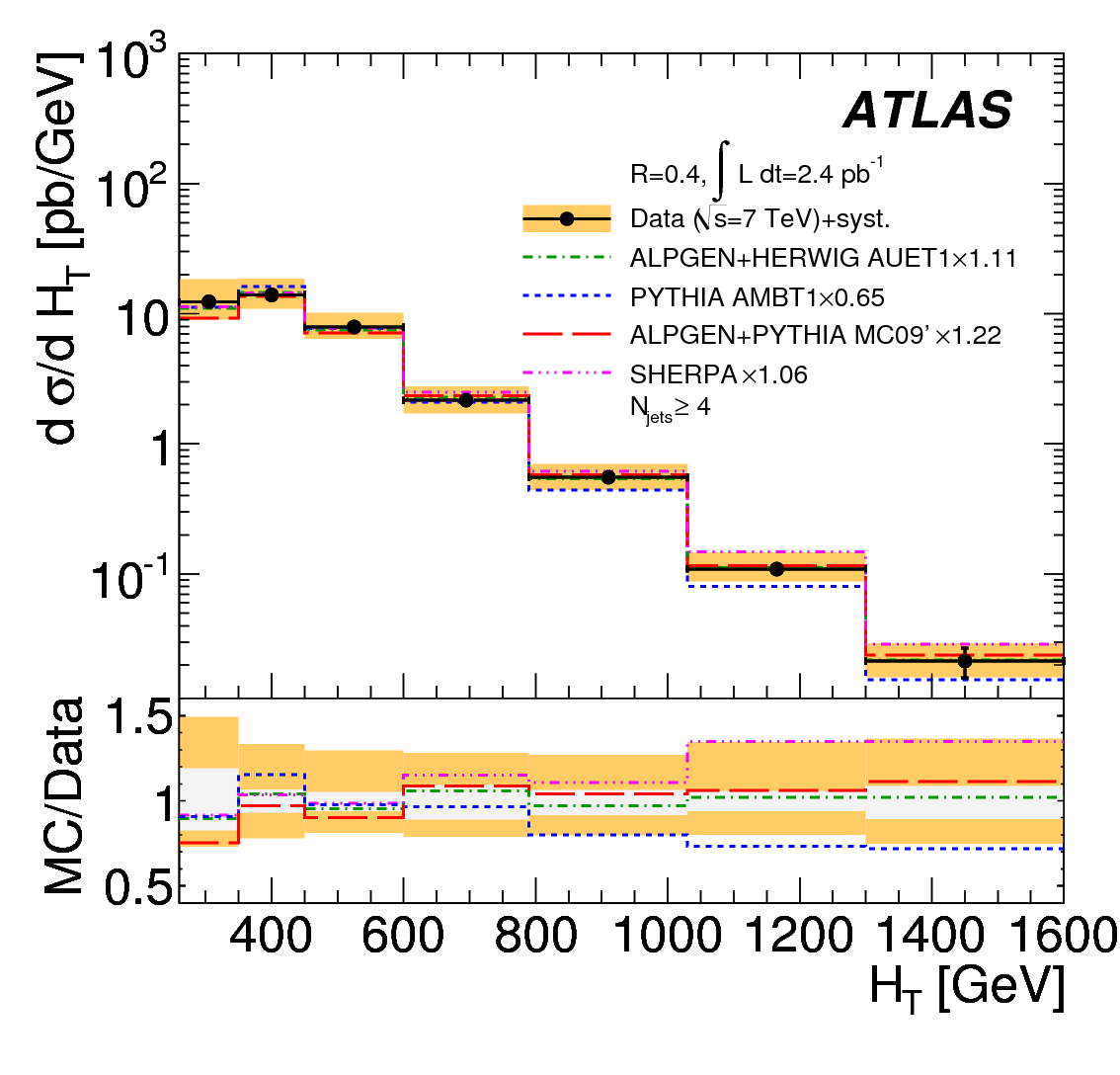}
\caption{Differential multijet cross section as a function of
  \HT~\protect\cite{ATL-2011-030}. Same as Figure~\protect\ref{fig:muljetpttj},
  but for or $\Njets\ge4$.
  \label{fig:muljetptvj}}
\end{figure}

 In addition to the multiplicity also the differential cross sections for
 multijet production as a function of \HT, and for different jet
 multiplicities is investigated.
 In this analysis, the systematic uncertainty is about (10-20)$\%$ across \pt,
 and increases to about 30$\%$ for the fourth leading jet differential
 cross section.
 The results for $\Njets\ge3$ and $\Njets\ge4$ are shown in
 Figure~\ref{fig:muljetpttj} and Figure~\ref{fig:muljetptvj}.
%
%
 Again, within uncertainties the data in both multiplicity bins can be described
 by all shown predictions.
 In these distributions the LO $2\to 2$ prediction from the \Pythia\ program is
 steeper than the LO $2\to n$ predictions from either the \Alpgen\ or the
 \Sherpa\ package, a trend that can generally be observed when comparing LO
 $2\to 2$ to LO $2\to n$ predictions.
%
%
 The shape differences in the two predictions based on the \Alpgen\ software,
 but using different programs and tunings for the soft part of the simulation
 implemented in \Herwig\ or \Pythia\ are very small, demonstrating a low
 sensitivity of the shape of these differential cross sections to soft effects.

 In addition to inclusive jet production, also more detailed investigations in
 quest for identifying BFKL signatures are performed~\cite{ATL-2011-029}.
 In this analysis, starting from a dijet system defining a rapidity gap, the
 properties of that gap are investigated.
 The selection requirements for jets obtained with $R=0.6$ are: $\arap<4.4$,
 $\pt>20$~\GeV, and an average transverse momentum of the two jets of the dijet
 system of $\ptb>50$~\GeV.
%
\begin{figure}[tbp!]
\centering
\includegraphics[width=0.49\textwidth]{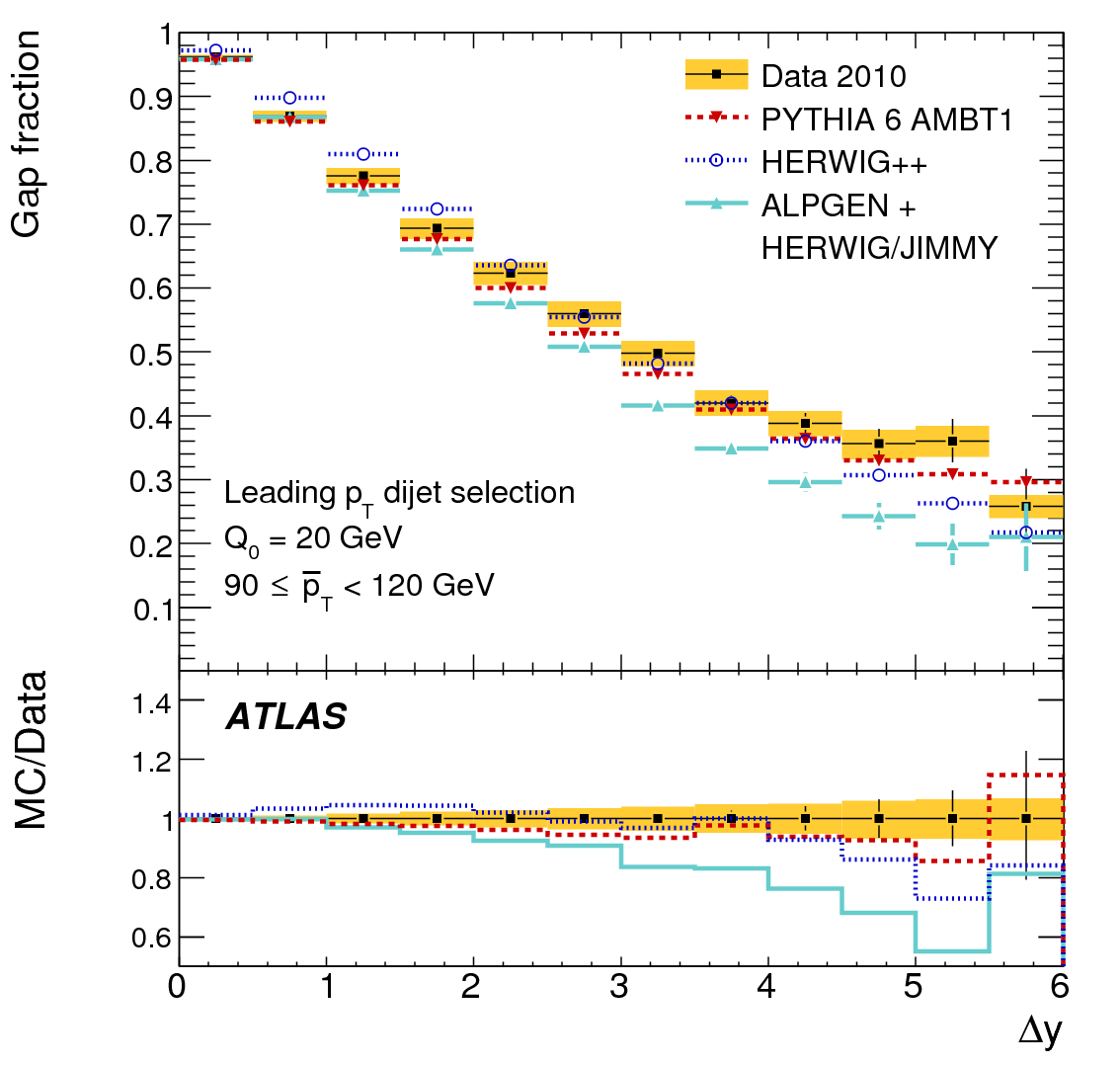}
\caption{The measured gap fraction as a function of \Dy\ for a bin in
  \ptb~\protect\cite{ATL-2011-029}. The vertical bars represent the statistical
  uncertainty, the band indicates the systematic uncertainty. The data are
  compared to a number of predictions detailed in the text.
  \label{fig:gapfrady}}
\end{figure}
%
%
\begin{figure}[tbp!]
\centering
\includegraphics[width=0.49\textwidth]{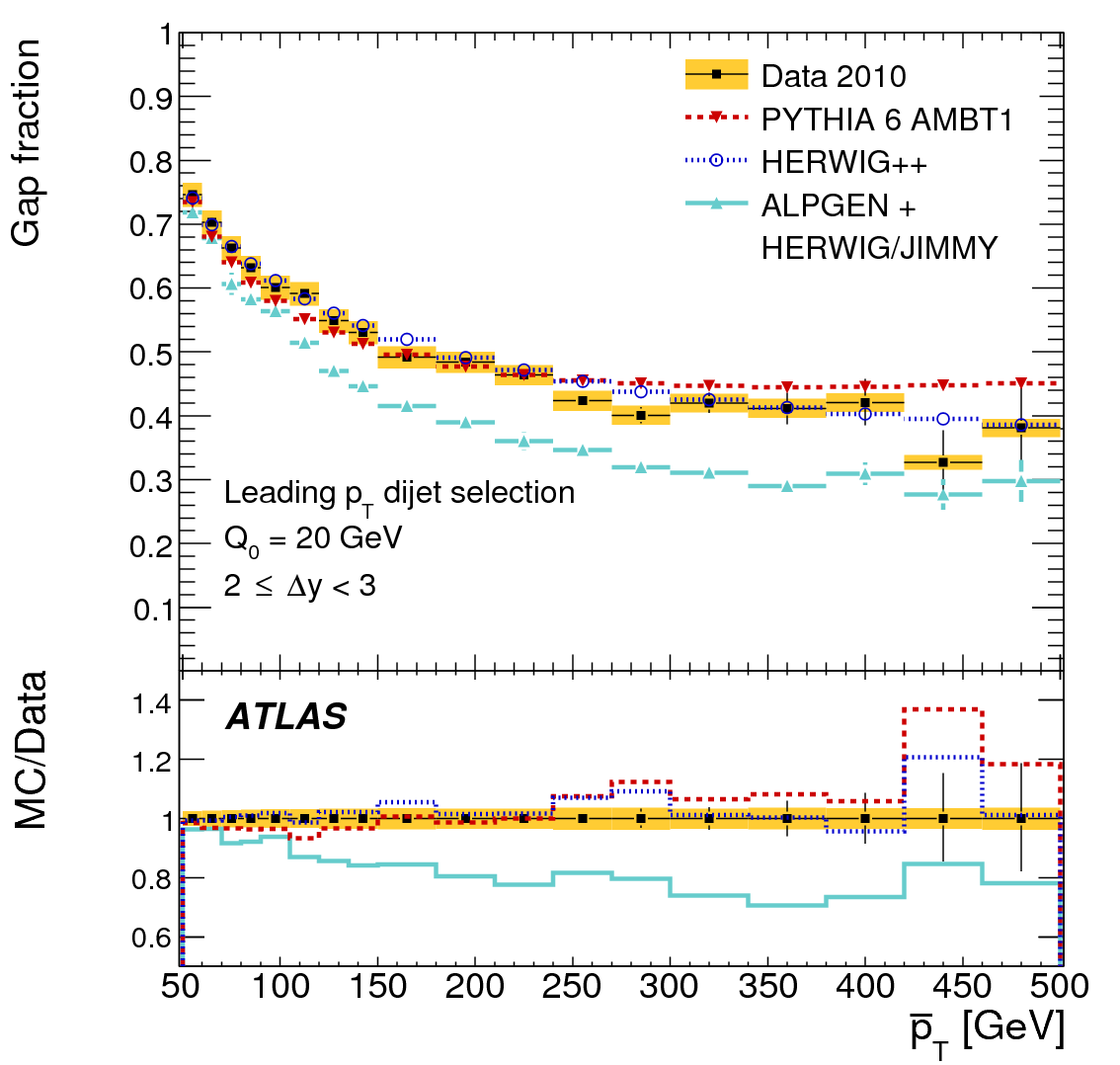}
\caption{The measured gap fraction as a function of \ptb\ for a bin in
  \Dy~\protect\cite{ATL-2011-029}. See Figure~\ref{fig:gapfrady} for details.
  \label{fig:gapfrapt}}
\end{figure}
%
 From these jets, the dijet system is either formed from the two highest
 \pt\ jets (leading \pt\ dijet selection), which typically have rather similar
 \pt, or from the two jets with the largest rapidity gap, for which typically
 their invariant mass is much larger than their \ptb.
 The gap properties investigated are either the gap fraction, i.e.~the fraction
 of events that do not contain any jet above a certain \pt\ threshold, chosen to
 be $\qn=20$~\GeV, i.e.~this scale satisfies $\qn\gg\Lambda$, or the average jet
 multiplicity of exactly those additional jets.
 These observables probe wide angle soft gluon radiation for $\qn\ll \ptb$, BFKL
 dynamics for large \Dy, and finally colour singlet exchange if both conditions
 are fulfilled at the same time.

 As an example, for the leading \pt\ dijet selection the corrected gap fraction
 is shown in Figure~\ref{fig:gapfrady} as a function of the rapidity gap
 \Dy\ for a given bin in \ptb, and in Figure~\ref{fig:gapfrapt} as a function of
 \ptb\ for a given bin in \Dy. The corrections to the stable particle level
 amount to about (2-4)$\%$.
 The JES uncertainty is about (2-5)$\%$ for the central region and 13$\%$ for
 the forward region, defined as $\arap>3.2$. The resulting uncertainty on the
 gap fraction is about 3$\%$~(7$\%$) for the same rapidity ranges.
 The comparison to the theoretical predictions reveals that the LO $2\to 2$
 predictions from \Pythia\ and \Herwig\ follow the data, except for large values
 of \Dy.  In contrast, the \Alpgen\ $2\to n$ model predicts too many jets,
 i.e.~a too small gap fraction, for both the \ptb\ and the \Dy\ dependence,
 except for low scales.

\begin{figure}[tbp!]
\centering
\includegraphics[width=0.49\textwidth]{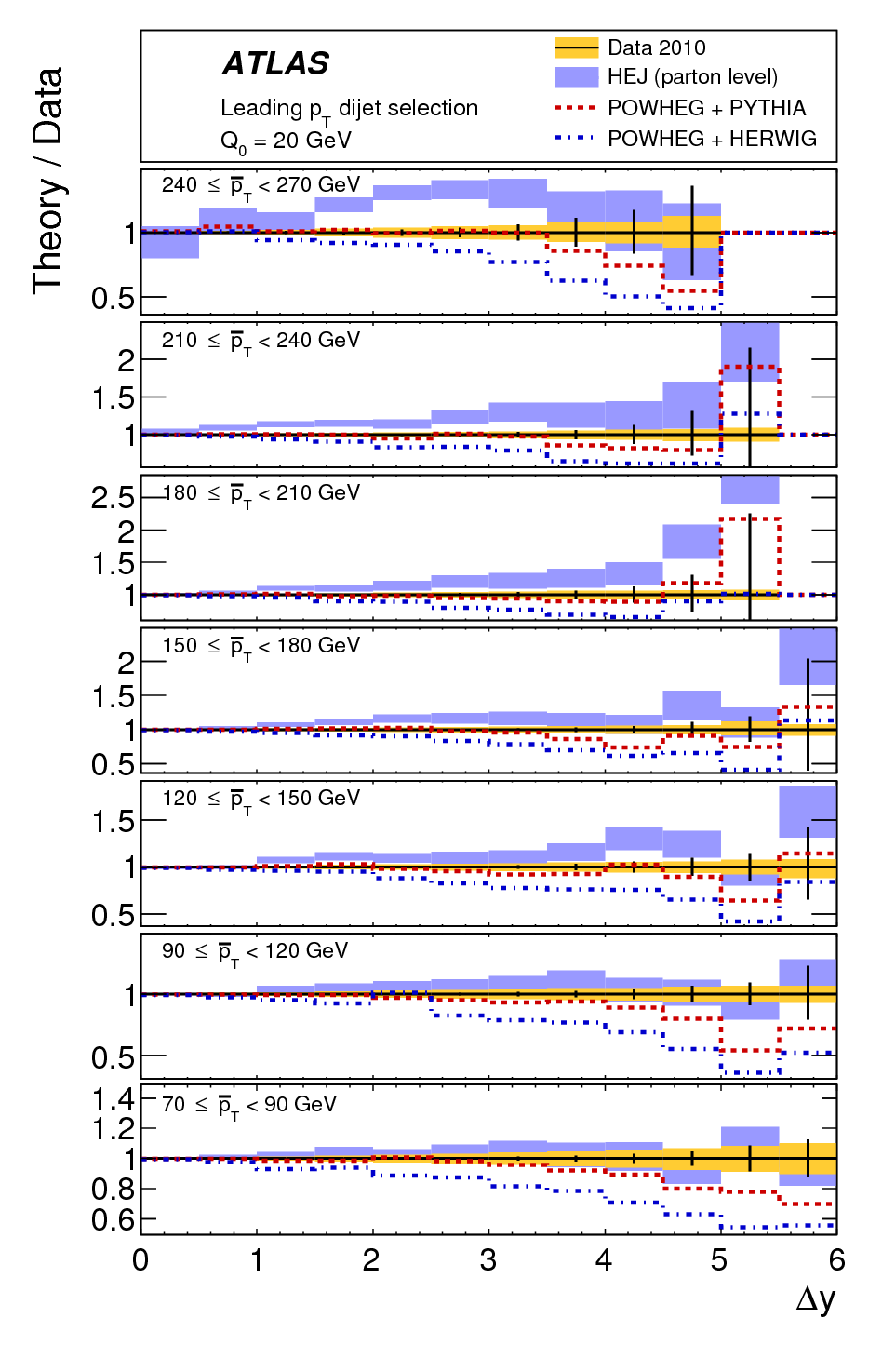}
\caption{The ratio of the predicted gap fraction to the one observed in the data
  as a function of \Dy, and for various bins in
  \ptb~\protect\cite{ATL-2011-029}. The data are displayed as in
  Figure~\protect\ref{fig:gapfrady}. The band not centred around unity
  represents the theoretical uncertainty in the \Hej\ calculation. For the two
  \Powheg\ predictions only the central result is shown.
  \label{fig:gapfraho}}
\end{figure}

 In Figure~\ref{fig:gapfraho} the ratio of the predicted and the observed gap
 fractions for various higher order predictions is displayed as a function of
 \Dy, and for a number of narrow ranges in \ptb.
 The NLO prediction from the \Powheg\ model generally has too much jet activity,
 with the \Pythia\ fragmentation being closer to the data than the one from
 \Herwig. 
 The partially large spread between \Powheg+\,\Pythia\ and
 \Powheg+\,\Herwig\ indicates regions of phase space with sizeable contributions
 from soft effects to this observable.
 The difference of the \Powheg\ model to the data increases for increasing \Dy.
 This can be attributed to the fact that the NLO plus PS prediction is lacking
 the full QCD ME contributions that become important as \Dy\ increases.

 The prediction from the \Hej\ program are shown at the parton level, i.e. they
 will in addition be subject to soft effects that may be sizeable, see the
 \Powheg\ discussion above.
 This prediction, apart from the region of low \Dy, has too few jets in the gap,
 especially so at large values of \Dy, and at large \ptb/\qn for all values of
 \Dy.
 Again, this deviation from the data is expected, since the theoretical QCD
 prediction implemented in the \Hej\ program is only a valid approximation in
 the limit where all jets have similar \pt.
%
%
%
\section{W/Z-Boson plus jet production}
\label{sec:wzboson}
 Jet production, together with an additional hard scale provided by the mass of
 a heavy boson, is investigated in the \WB\ plus 1-jet and \ZB\ plus 1-jet
 production processes, using the leptonic decays of the heavy bosons
 $W\to\ell\nu_\ell$ and $Z\to\ell^+\ell^-$ with $\ell=$ electron or
 muon~\cite{ATL-2011-041}.

 The driving idea in this analysis is to construct an observable with very small
 experimental uncertainty to perform a precise QCD test. Therefore, firstly, not
 individual cross sections, but the ratio of the \WB\ plus 1-jet and \ZB\ plus
 1-jet production cross sections is utilised, and secondly, this ratio is
 investigated as a function of the \pt\ threshold, and not in bins of \pt.

 In this analysis, jets have to fulfill $\pt>30$~\GeV and $\aeta<2.8$, and
 events with additional jets with $\pt>30$~\GeV are vetoed.
 The background contribution is small, below 5$\%$ in all channels, but for the
 QCD multijet background in the $W\to\mathrm{e}\nu_\mathrm{e}$ channel which is
 19$\%$.
 All background estimates are taken from Standard Model Monte Carlo samples,
 except for the QCD multijet background. This is because this background can at
 present not be reliably modelled in the Monte Carlo programs, and is therefore,
 as for most analyses, taken from sideband regions in the data.
%
\begin{figure}[tbp!]
\centering
\includegraphics[width=0.48\textwidth]{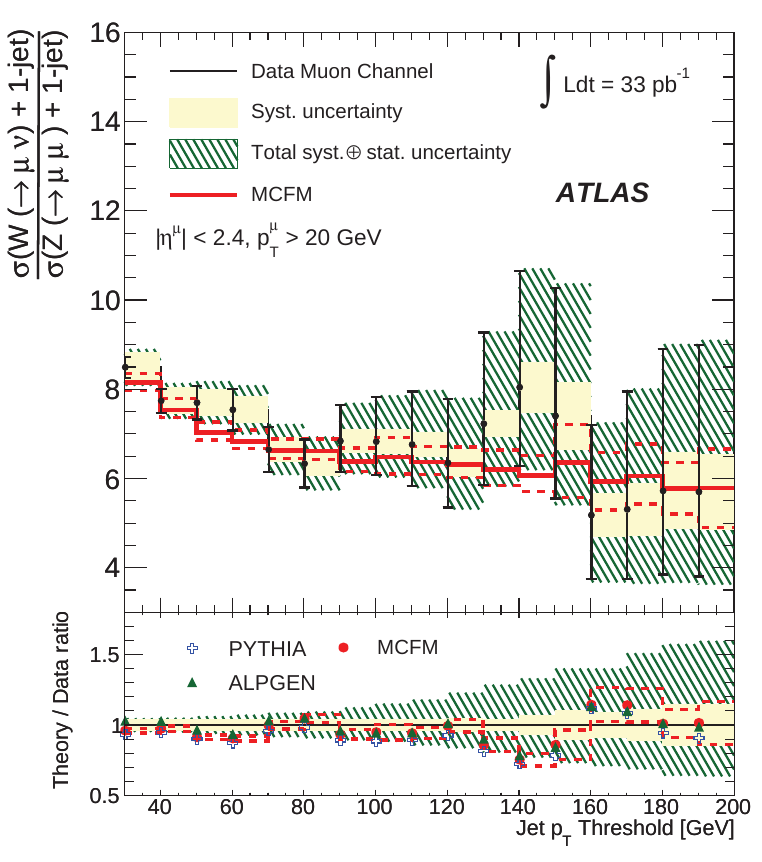}
\caption{The ratio of \WB\ plus 1-jet to \ZB\ plus 1-jet production in the muon
  decay channel as a function of the jet
  \pt\ threshold~\protect\cite{ATL-2011-041}. The data are shown as points at
  the respective threshold, together with their statistical uncertainty
  (vertical bars), their systematic uncertainty (inner band), and their
  total uncertainty (outer band). The dashed lines indicate the theoretical
  uncertainty on the MCFM prediction dominated by the PDF and scale
  uncertainties.
  \label{fig:wbopromu}}
\end{figure}
%

 The result in the muon decay channel, corrected to the stable particle level
 with \Pythia, is shown in Figure~\ref{fig:wbopromu}.
 As expected, the ratio decreases with increasing jet \pt\ threshold, because
 the effective scale of the interaction becomes large with respect to the
 difference in the heavy boson masses.
 The systematic uncertainty (inner band) on the ratio is about (5-10)$\%$, and
 in itself it has a large statistical component. 
 The largest contribution to the systematic uncertainty is due to imperfections
 in the heavy boson reconstruction, including effects from the lepton trigger,
 reconstruction and identification efficiencies, and scale uncertainties for the
 lepton and missing transverse energy measurements.
 For transverse momenta larger than about 50~\GeV, the total uncertainty is
 dominated by the statistical uncertainty (vertical bars).
 This uncertainty will soon decrease given the large statistics of the completed
 2011 LHC run, which amounts to about 5/fb.
%
\begin{figure}[tbp!]
\centering
\includegraphics[width=0.48\textwidth]{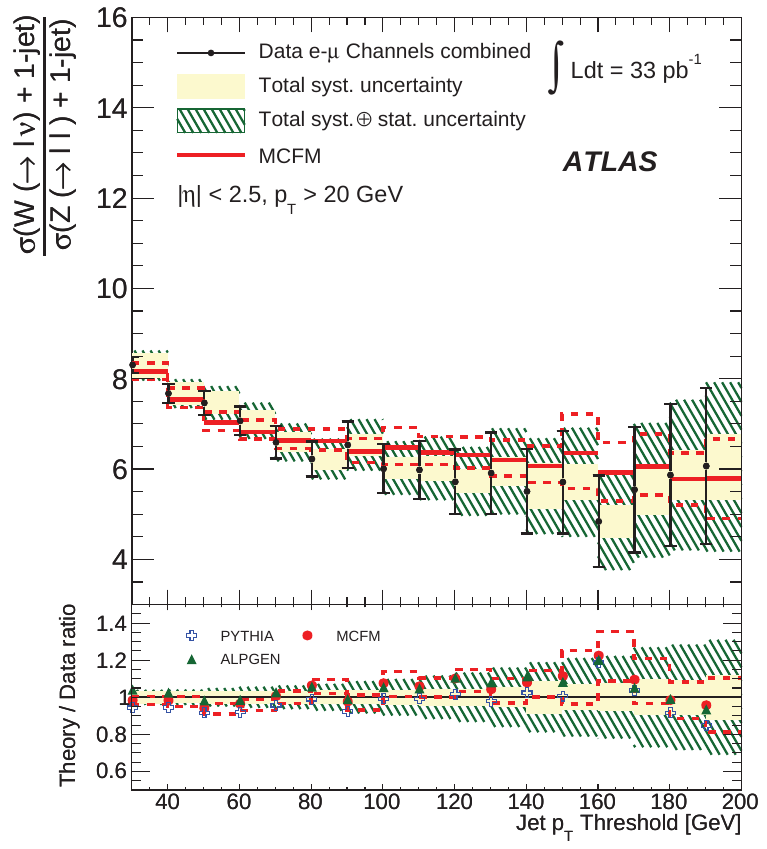}
\caption{Same as Figure~\ref{fig:wbopromu} but for both lepton decay channels
  combined~\protect\cite{ATL-2011-041}.
  \label{fig:wbopro}}
\end{figure}

 The measurements in both lepton decay channels are consistent and the combined
 result, evaluated for a common phase space region for the leptons, and based on
 data for an integrated luminosity of 33/pb, is shown in
 Figure~\ref{fig:wbopro}.
 The final value of the ratio, corrected to the phase space of the leptons
 indicated, and for the lowest \pt\ threshold of 30~\GeV, is
 \XZ{8.23}{0.18}{0.28}.

 The data are compared to three predictions: a LO $2\to 2$ prediction based on
 the \Pythia\ program; a LO $2\to n$ prediction from the \Alpgen\ software; and
 finally, an NLO ME calculation for $2\to W/Z\,+\,2$ partons based on the
 \Mcfm\ program.
 All predictions fall within the still large experimental uncertainty band, but
 clearly the deviations between data and predictions partly exceed the
 systematic uncertainty of the data.
 In particular, the data are well described by the NLO \Mcfm\ prediction, for
 which the uncertainty (shown as dashed lines) is driven by the PDF uncertainty
 and the one due to scale variations.
 The experimental systematic uncertainties are smaller than those of these
 predictions, especially at large \pt\ thresholds such that, after including the
 2011 data, the experimental precision will challenge the NLO theoretical
 prediction.
%
%
\section{Top-quark pair production}
\label{sec:ttbar}
 The LHC is a top quark factory. At the present proton-proton centre-of-mass
 energy of $\rts=7$~\TeV, the theoretical value of the \ttbar\ production
 cross section, obtained from a computation approximating the NNLO prediction,
 and for an assumed input top quark mass of $\mt=172.5$~\GeV\ is about
 \sigttbar, with an uncertainty of about \dsigttbar~\cite{ALI-1101}, see below
 for details.
 This cross section is about 20 times larger than the corresponding
 cross section at the Tevatron.
 The LHC experiments have already analysed a wide spectrum of top quark 
 physics.
 The two observables discussed are the \ttbar\ production
 cross section~\cite{ATL-2010-016,CMS-1107} and the top quark mass
 \mt~\cite{CMS-1101}.
 
 The two most important quantities to be precisely evaluated for a cross section
 determination are: the selection efficiency for signal events, and the amount
 of background events present in the data.
 To achieve a high precision estimate in the LHC analyses, not only the overall
 normalisation of the background is used, but the shape of its contribution as
 a function of one or more variables discriminating signal and background
 processes is utilised.

\begin{figure}[tbp!]
\centering
\includegraphics[width=0.47\textwidth]{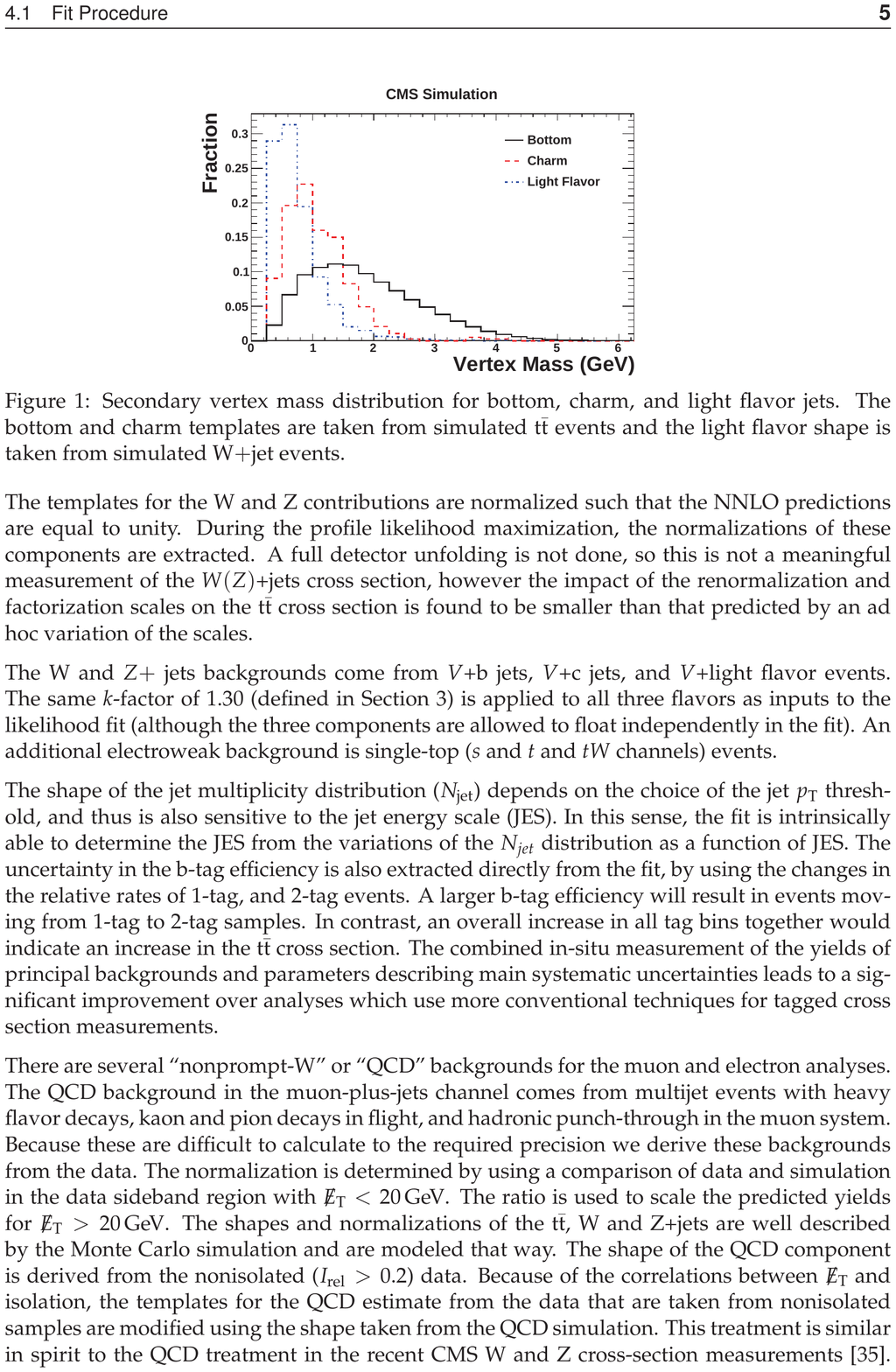}
\caption{The \ttbar\ production cross section~\protect\cite{CMS-1107}. Shown is
  the vertex mass obtained from the charged particles assigned to a secondary
  vertex as predicted in simulation.
  \label{fig:ttbpro}}
\end{figure}

 An example of such a variable, the mass distribution of identified secondary
 vertices, is shown in Figure~\ref{fig:ttbpro}~\cite{CMS-1107}. This figure
 shows that the mass distributions originating from either light-quark jets or
 \bquark\ jets are significantly different.
 The mass distribution from jets stemming from charm quarks falls in between the
 two.
 Using this discriminant variable, a number of statistically independent
 sub-sets of data with different signal to background compositions are
 exploited.
%
%

\begin{figure}[tbp!]
\centering
\includegraphics[width=0.48\textwidth]{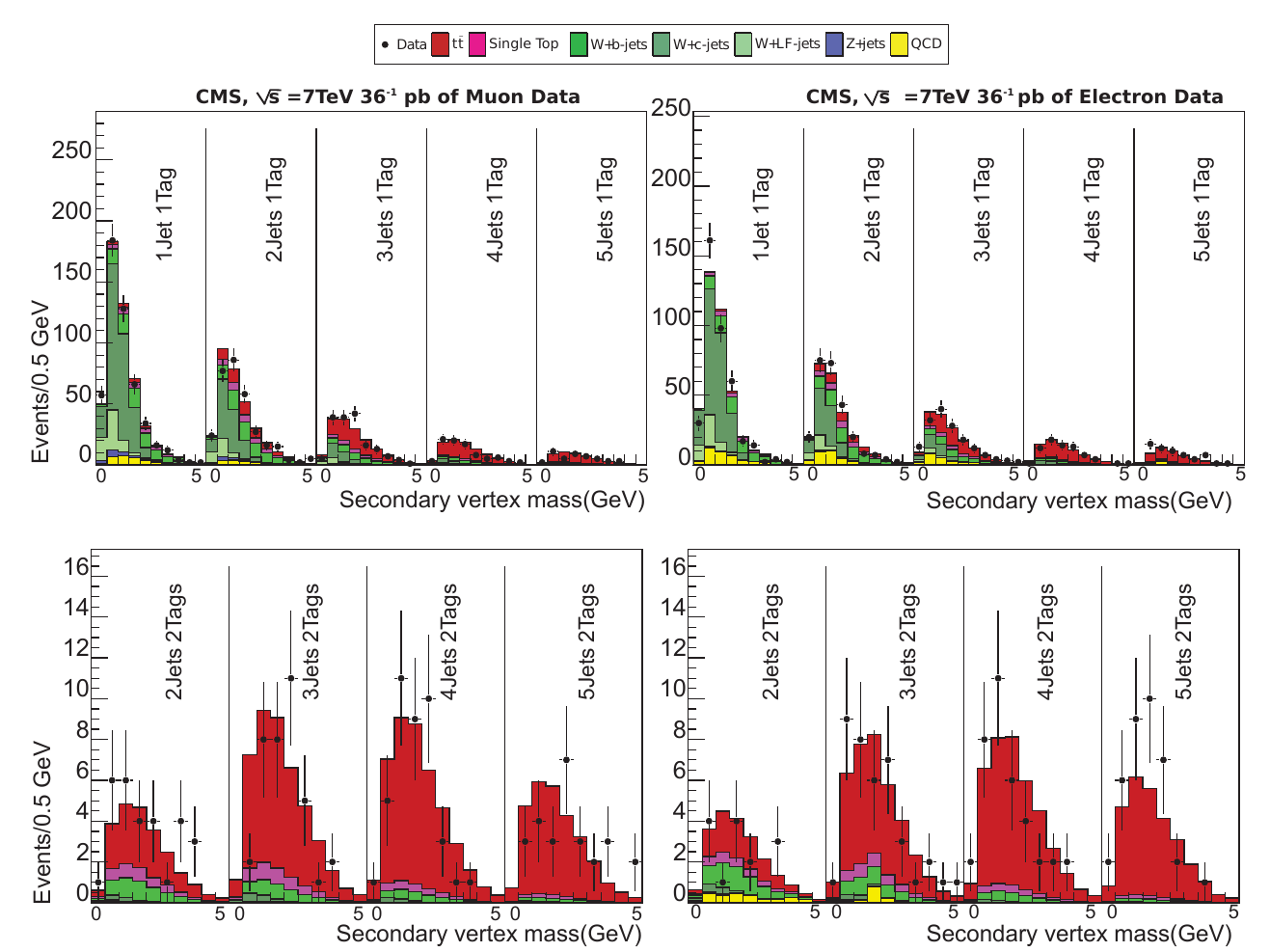}
\caption{The \ttbar\ production cross section~\protect\cite{CMS-1107}.  Shown
  are the mass distributions of identified secondary vertices for the two lepton
  decay channels, and for different jet- and \bjet\ multiplicities. The data are
  shown with their statistical uncertainty, together with the fitted
  contributions of the predicted signal and background samples.
  \label{fig:ttbprofit}}
\end{figure}

 Figure~\ref{fig:ttbprofit} shows this set of distributions used in the fit for
 the \ttbarlj\ decay in the \ejets\ and \mjets\ channels, for different jet
 multiplicities, and depending on the number of \btagged\ jets.
 The simulated signal events accumulate at large vertex masses, and their
 fraction grows with the numbers of observed jets and \bjets.

 The use of the profile likelihood method allows systematic uncertainties, which
 are treated as nuisance parameters in the fit, to cancel each other within
 bounds. Therefore, this method in general leads to smaller uncertainties than
 are achieved when individually varying systematic effects to ascertain the
 corresponding uncertainties. 
 This method requires a very good modelling of the correlation of the systematic
 uncertainties, since otherwise fortuitous cancellations can happen.
%
%

 The variations of the systematic uncertainties are individually constrained in
 the fit by Gaussian priors. The different sources of systematic uncertainty are
 correlated or anti-correlated in the fit by up to absolute 70$\%$.

 The determination of the cross section in the \ejets\ and \mjets\ channels
 leads to consistent results.
 The measured cross section, obtained from simultaneously fitting the
 distributions of both channels using data corresponding to only 36/pb of
 luminosity, Figure~\ref{fig:ttbprofit}, has an uncertainty of about 20~pb and
 is already limited by systematic effects.
 The measured cross section value is $\ttbcross=(\XZ{154}{9}{17})$~pb with an
 additional uncertainty of 6~pb due to the uncertainty on the
 luminosity~\cite{CMS-1107}.
%
\begin{figure}[tbp!]
\centering
\includegraphics[width=0.48\textwidth]{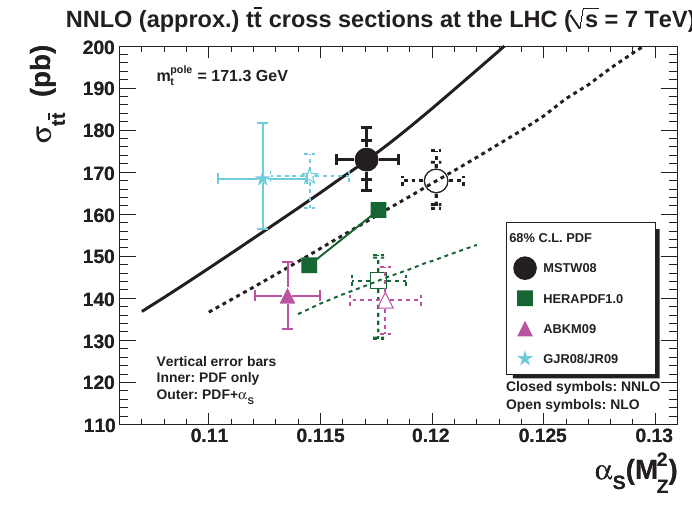}
\caption{The predicted \ttbar\ cross section at the LHC, obtained from an NLO
  (open markers) and an approximate NNLO (closed markers) calculation, as a
  function of the strong coupling constant. The values are given for
  $\rts=7$~\TeV\ and for a number of PDF sets~\cite{WAT-1101}. The markers are
  placed at the predicted cross section and the \almzq\ value of the respective
  PDF set. The horizontal bar span the \almzq\ uncertainty, and the vertical
  bars indicate the PDF uncertainty of the cross section (inner bar), and the
  PDF and \al\ uncertainty (outer bar). The lines indicate the cross section
  variation with the \al\ dependent additional PDF sets.
  \label{fig:sigtop}}
\end{figure}
%
 
\begin{figure}[tbp!]
\centering
\includegraphics[width=0.48\textwidth]{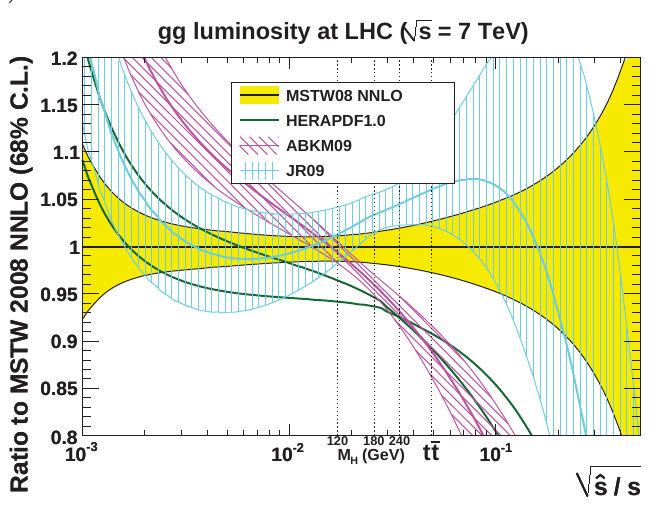}
\caption{The NNLO gluon-gluon luminosity at the LHC for $\rts=7$~\TeV\ as a
  function of the gluon-gluon invariant mass scaled to the centre-of-mass
  energy, as predicted by a number of PDF sets~\cite{THO-1101}. The
  distributions are all normalised to the NNLO prediction from MSTW 2008. The
  bands indicate the uncertainties obtained from the additional pdf sets.
  \label{fig:gglum}}
\end{figure}

 This measured cross section can be used to further constrain the PDFs. Since,
 at the LHC the \ttbar\ production process is largely dominated by gluon-gluon
 processes, in contrast to the Tevatron, where it is dominated by
 quark-antiquark processes, this mostly concerns the gluon distribution
 function.
 Figure~\ref{fig:sigtop} shows the present theoretical knowledge of this
 cross section based on an NLO and an approximate NNLO calculation, and for
 various PDF sets~\cite{WAT-1101}. The points are given at the respective value
 of \almzq\ of the corresponding PDF set. The NNLO corrections are small.
 The smallest cross section values are predicted when using the ABKM09 PDF
 set. This is correlated to the smallest predicted gluon-gluon luminosity at
 the \ttbar\ production threshold and above, as shown in
 Figure~\ref{fig:gglum}~\cite{WAT-1101}.
%
%
 The difference in production cross section between the highest and lowest
 prediction at NNLO is about 33~pb, i.e. about 1.5$\sigma$ of the above
 described measurement.
 With improved measurements that are underway, and by combining the results from
 the ATLAS and CMS experiments, the gluon PDF can be significantly constrained.

 The present average value of the top quark mass of $\mt=(\mttevo)$~\GeV is
 obtained from direct measurements performed at the Tevatron~\cite{TEV-1101},
 and has a total uncertainty of $0.6\%$.
 The main methodology used to determine \mt\ at hadron colliders consists of
 measuring the invariant mass of the decay products of the top quark candidates
 and deducing \mt\ using sophisticated analysis methods.
 The most precise measurements of this type use the decay \ttbarlj\ with
 $\mathrm{lepton}=e,\mu$, where one \WB\ decays into a lepton and a neutrino and
 the other into a pair of quarks.
 These measurements rely on Monte Carlo programs to simulate the \ttbar\ final
 state. The experimental observables are constructed such that they are unbiased
 estimators of the top quark mass used as an input parameter in the Monte Carlo,
 denoted with \mtMC, which is verified using pseudo-experiments performed on
 large scale Monte Carlo simulated event samples.
 Consequently, the top quark mass determined this way corresponds to \mtMC.

 On the theoretical side, there are a number of definitions of the mass. The
 definition of the pole mass, \mtpole, basically regards the quark as free and
 long lived.
 In contrast, for the \msb\ mass definition, \mtmsb, the mass is treated like a
 coupling.
 The masses expressed in the two renormalisation schemes are related, and
 consequently can be converted into one another. 
 Their difference is sizable compared to the experimental precision:
 $\mtpole=172$~\GeV\ leads to approximately $\mtmsb=162$~\GeV, a difference of
 about 6$\%$, i.e.~ten times the experimental uncertainty.
 Non of these definitions coincides with \mtMC\ defined above, which leads to a
 problem in interpreting the experimental results.
 There are theoretical arguments~\cite{BUC-1101} suggesting that \mtpole\ is
 closer to \mtMC\ than \mtmsb, and that \mtpole\ is expected to be
 \Ord{1~\mbox{GeV}} larger than \mtMC, but no proof of this relation from first
 principles exists.

 Theoretically, the \ttbar\ pair production cross section \ttbcrossv{\mtpole} is
 known in a given renormalisation scheme. The calculations are performed at NLO,
 NLO+(N)NLL or approximate NNLO precision, and have a strong dependence on the
 top quark mass.
 Consequently, it was suggested that by utilising this dependence, and
 extracting the top quark mass from the cross section, the problem explained
 above is absent, i.e.~the resulting \mt\ corresponds to a mass in a
 theoretically well defined concept.
 In addition, when the top quark mass is extracted from a comparison of the
 measured production cross section with its prediction as a function of the
 mass, one could profit from the fact that the relative uncertainty of the cross
 section translates into an about five times smaller uncertainty on the top
 quark mass, when neglecting the theoretical uncertainties.
%
\begin{figure}[tbp!]
\centering
\includegraphics[width=0.48\textwidth]{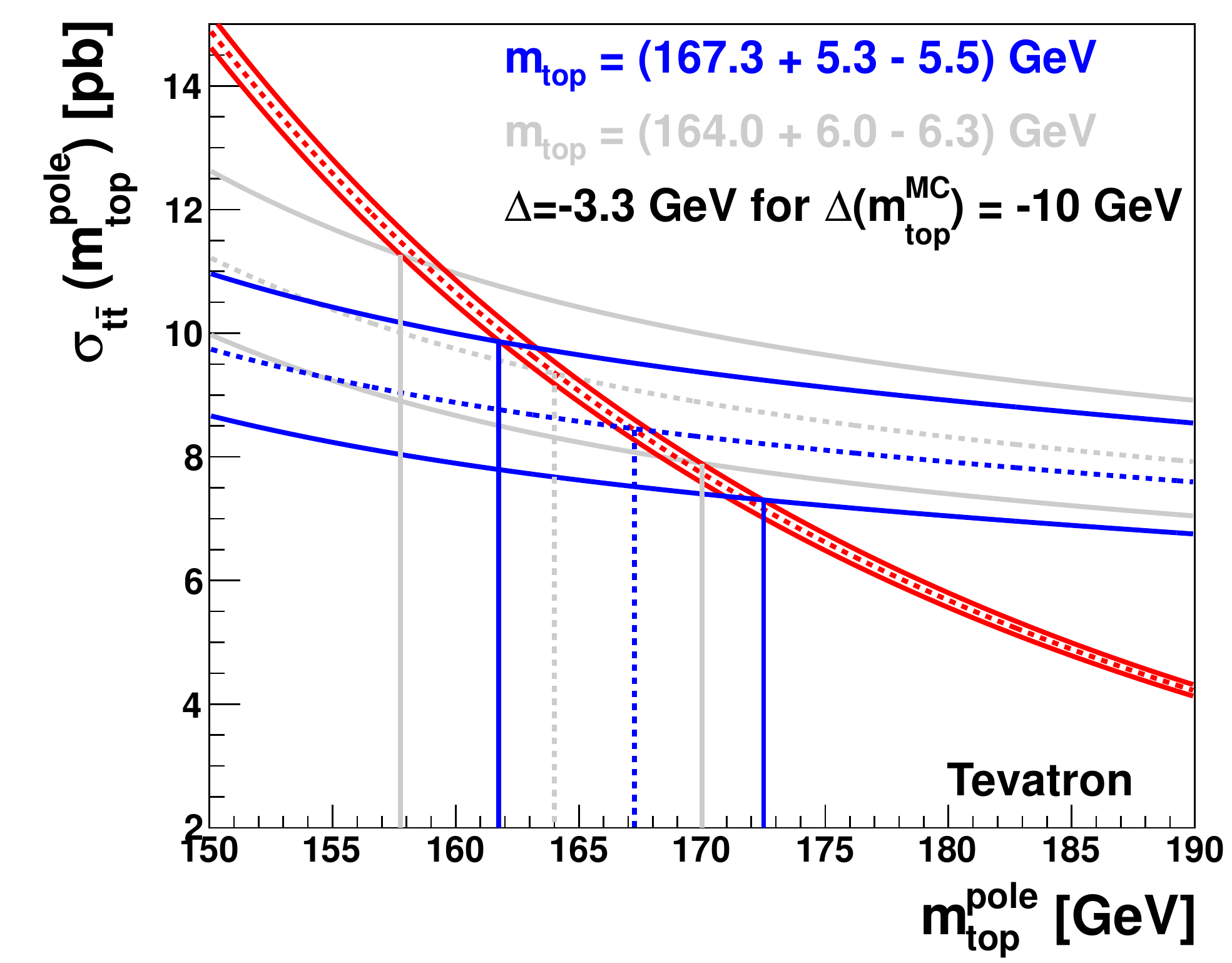}
\caption{Dependence of the \ttbar\ production cross section on the top quark
  mass \mt\ for Tevatron conditions.
  \label{fig:mtoptev}}
\end{figure}
%

 This concept is only valid if the experimental determination of \ttbcross\ does
 not depend on the value of \mt\ itself, which unfortunately is not the case.
 This is because also for the measurement of the \ttbar\ cross section,
 \mtMC\ is needed, since the Monte Carlo models are indispensable for evaluating
 the acceptance, efficiency and the systematic uncertainties in the
 experimental determination of the cross section.

 Figure~\ref{fig:mtoptev} shows the dependence of the measured production cross
 section~\cite{D00-1101} and its experimental uncertainty as a function of
 \mtpole\ for $\mtMC=\mtpole$ (flattest of the three bands), for the Tevatron
 conditions.
 The measured cross section value is $\ttbcross=(\Y{8.13}{1.02}{0.90})$~pb,
 with a symmetrised uncertainty of about 12$\%$.
%
\begin{figure}[tbp!]
\centering
\includegraphics[width=0.48\textwidth]{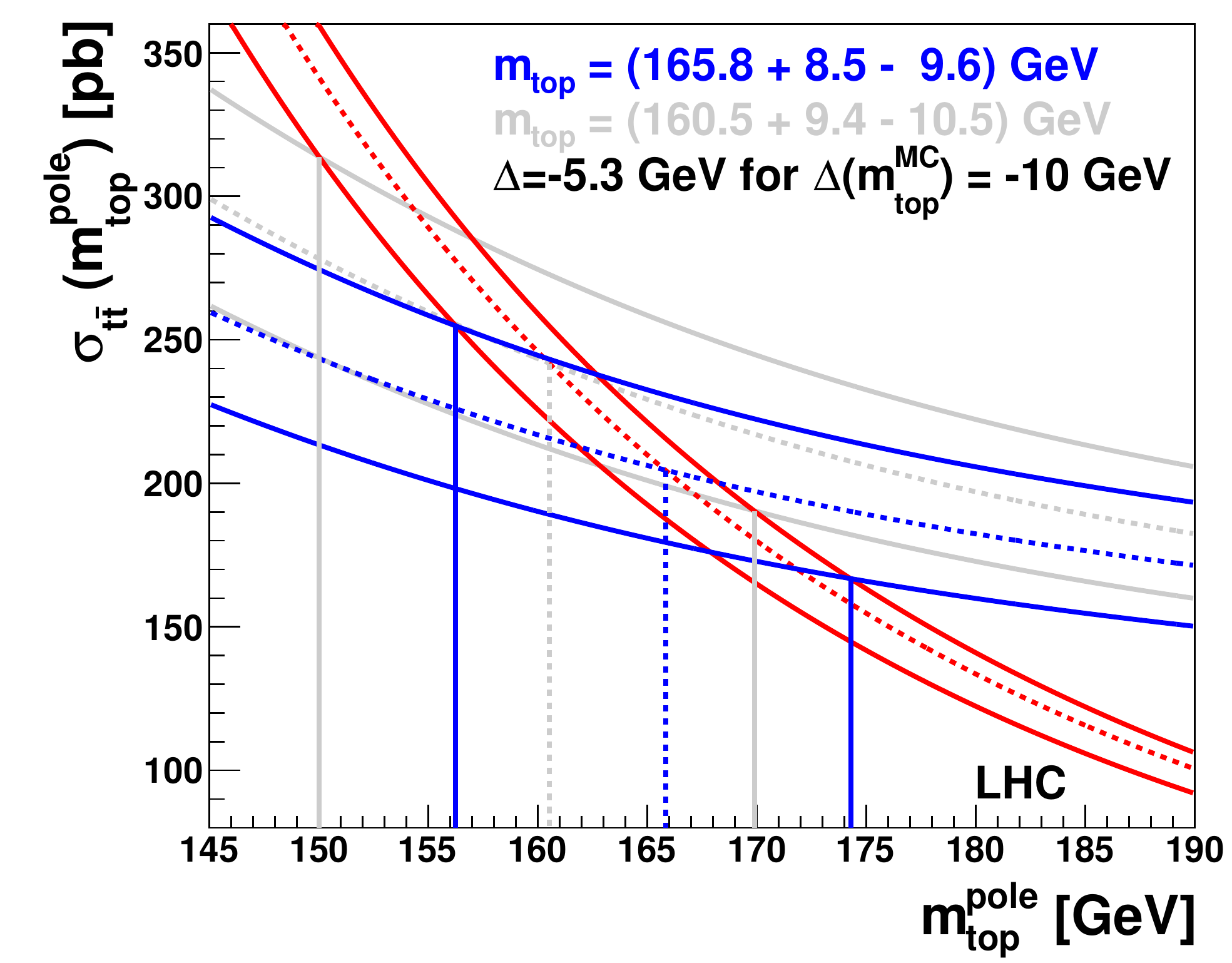}
  \caption{Same as Figure~\protect\ref{fig:mtoptev}, but simulating the LHC
    conditions at $\rts=7$~\TeV.
    \label{fig:mtoplhc}}
\end{figure}
%
 Since the measured cross section depends on the value of \mtMC, its relation to
 \mtpole\ is important as can be seen from the second shifted band shown for the
 assumption $\mtMC=\mtpole-10$~\GeV, which now means \mtMC\ approximately
 coincides with \mtmsb.

 Experimentally, the top quark mass is essentially extracted from the overlap of
 the experimental band and the theoretical band (the steepest band shown), which
 in this figure is based on~\cite{LAN-0901}. The result is indicated by the
 vertical lines in the figure.
 For a given assumption on \mtMC, this yields \mt\ with an uncertainty of about
 3$\%$, showing the aforementioned reduction in relative uncertainty.
 However, the uncertainty of where to put the experimental band leads to an
 additional uncertainty on \mt.
 For the example of the two extreme assumptions made in
 Figure~\ref{fig:mtoptev}, the corresponding difference in the extracted top
 quark mass is about 3~\GeV.

 A similar situation is shown in Figure~\ref{fig:mtoplhc}, again using the
 predictions from~\cite{LAN-0901}, but this time for LHC running conditions and
 assuming a somewhat steeper dependence of the measured cross section on \mtMC.
 For this situation, the difference in the extracted top quark mass is
 correspondingly larger, and amounts to about 5~\GeV.

 This investigation shows that to mitigate this uncertainty it is most important
 to find an \mt\ independent selection, i.e.~to select the signal events while
 depending as little as possible on absolute energy scales which directly relate
 to the actual value of \mt.
 In addition, to reach a precision on \mt\ of 0.6$\%$ as obtained in the direct
 measurements, this indirect extraction needs to achieve about a 3$\%$ precision
 on the measured cross section, which is a big challenge.
%
%
\section{Conclusions}
\label{sec:concl}
 The analyses of the data from the first year of LHC running at a proton-proton
 centre-of-mass energy of $\rts=7$~\TeV\ resulted in a large variety of physics
 results concerning QCD observables, only a small part of which could be
 discussed here.

 The investigation of jet production for a number of jet multiplicities already
 proved a helpful tool to better constrain QCD predictions of various types
 implemented in a large number of programs.
 The production of a heavy gauge bosons in conjunction with jets constitutes a
 high precision QCD test, the potential of which has just been started to be
 explored.

 The LHC is a top quark factory with a twenty times larger production cross
 section for pair production of top quarks than at the Tevatron.  The first
 measurements already give interesting hints on the size of the gluon PDF at
 large values of x.
 The determination of the top quark mass poses interesting challenges to the
 experiments and also to the interpretation of the measured values.

 All results presented were based on about 35/pb of data from the 2010 LHC
 run. By now about 5/fb of data each have been collected by the ATLAS and CMS
 experiments in 2011.
 With this huge amount of data, the statistically limited analyses from 2010
 data, can now be considerably expanded. For the analyses discussed in this
 paper, this especially applies to the measurements of the ratio of the heavy
 gauge bosons plus 1-jet cross sections.

 The constantly increasing specific luminosity leads to more and more
 proton-proton interactions per bunch crossing which pose an increasing
 challenge to the proper treatment of the pileup.
 With an even better understanding of the detectors, a large number of
 interesting and precise measurements are ahead of us. The LHC experiments will
 constantly extend the highest scales at which QCD has ever been probed at
 accelerators.
%
%
\section*{Acknowledgements}
 I like to thank the organisers not only for the stimulating spirit of this
 meeting, but even more for their tireless enthusiasm during all the passed
 years. I am glad I could participate in the last conference of this series.
%
%
\nocite{*}
\bibliographystyle{elsarticle-num}
\bibliography{nisius.bib}
%
%
\end{document}